\documentclass[11pt]{article}	% Req for JHEP
\usepackage{color} 
\usepackage{amsmath}
\usepackage{amsfonts}
\usepackage{amssymb}
\usepackage{float}
\usepackage{caption}
\usepackage{graphicx}
\usepackage{slashed}            % for slashed characters in math mode
\usepackage{subfig}             % for sub figures
\usepackage{bbm}                % for \mathbbm{1} (unit matrix)
\usepackage{xspace}				% For spacing after commands
\usepackage{collref}				% for grouping BiBTeX entries
\usepackage{framed}				% for framed text
\usepackage{placeins}
\usepackage{jheppub}							% JHEP
\usepackage{mathrsfs}
\usepackage{tikz}
\usetikzlibrary{arrows,shapes}
\usetikzlibrary{trees}
\usetikzlibrary{matrix,arrows} 				% For commutative diagram
\usetikzlibrary{positioning}				% For "above of=" commands
\usetikzlibrary{calc,through}				% For coordinates
\usetikzlibrary{decorations.pathreplacing}  % For curly braces
\usepackage{pgffor}							% For repeating patterns

\usetikzlibrary{decorations.pathmorphing}	% For Feynman Diagrams
\usetikzlibrary{decorations.markings}
\usetikzlibrary{snakes}

\tikzset{
    vector/.style={decorate, decoration={snake}, draw},
	provector/.style={decorate, decoration={snake,amplitude=2.5pt}, draw},
	antivector/.style={decorate, decoration={snake,amplitude=-2.5pt}, draw},
    fermion/.style={draw, postaction={decorate},
        decoration={markings,mark=at position .55 with {\arrow[draw]{>}}}},
    fermionbar/.style={draw, postaction={decorate},
        decoration={markings,mark=at position .55 with {\arrow[draw=black]{<}}}},
    fermionnoarrow/.style={draw},
    gluon/.style={decorate, draw,decoration={coil,amplitude=4pt, segment length=6pt}, line width=1},
    scalar/.style={dashed,draw, postaction={decorate},
        decoration={markings,mark=at position .55 with {\arrow[draw]{>}}}},
    scalarbar/.style={dashed,draw, postaction={decorate},
        decoration={markings,mark=at position .55 with {\arrow[draw]{<}}}},
    scalarnoarrow/.style={dash pattern = on 6 pt off 3 pt,draw},
    electron/.style={draw, postaction={decorate},
        decoration={markings,mark=at position .55 with {\arrow[draw]{>}}}},
	bigvector/.style={decorate, decoration={snake,amplitude=4pt}, draw},
	vectorscalar/.style={loosely dotted,draw, postaction={decorate}},
}
\RequirePackage{xspace}
\usepackage{relsize}
\usepackage{slashed}

\def\lsim{\mathrel{\rlap{\lower4pt\hbox{\hskip1pt$\sim$}}
    \raise1pt\hbox{$<$}}}                % less than or approx. symbol
\def\gsim{\mathrel{\rlap{\lower4pt\hbox{\hskip1pt$\sim$}}
    \raise1pt\hbox{$>$}}}                % greater than or approx. symbol
\renewcommand{\tilde}{\widetilde} % dinky tildes look silly

\usepackage{pstricks} 	% Please don't use this

\usepackage{hyperref}   % Doesn't always play well with other packages
\usepackage{cancel}
\newcommand{\be}{\begin{eqnarray}}
\newcommand{\ee}{\end{eqnarray}}

\title{\boldmath Soft RPV Through the Baryon Portal}

\author[a]{Gordan Krnjaic} 
\author{and} 
\author[b]{Yuhsin Tsai}

\affiliation[a]{\it Perimeter Institute for Theoretical Physics   \\ 
\it  Waterloo, Ontario, Canada}

\affiliation[b]{\it Physics  Department, University of California Davis  \\ 
\it Davis, California, USA}

\emailAdd{gkrnjaic@perimeterinstitute.ca}
\emailAdd{yhtsai@ucdavis.edu}

\abstract{Supersymmetric (SUSY) models with $R$-parity generically 
predict sparticle  decays with invisible neutralinos, which yield distinctive missing energy events at colliders. 
Since most LHC searches are designed with this expectation, 
 the putative bounds on sparticle masses become considerably weaker if $R$-parity is violated 
 so that squarks and gluinos decay to jets with large QCD backgrounds.  
Here we introduce a  scenario in which baryonic $R$-parity violation (RPV) arises effectively from soft
 SUSY breaking interactions, but leptonic RPV remains accidentally forbidden to 
  evade constraints from proton decay and FCNCs. 
  The model features a global $R$-symmetry that initially 
 forbids RPV interactions, a hidden 
 $R$-breaking sector, and a heavy mediator that communicates this breaking to the visible sector.  
After $R$-symmetry breaking, the mediator is integrated out and an effective RPV $A$-term arises at tree 
level; RPV couplings between quarks and squarks arise only at loop level and receive additional suppression. Although this 
mediator must be heavy compared to soft masses, the model introduces no new hierarchy since 
viable RPV can arise when the mediator mass is near the SUSY breaking scale.
In generic regions of parameter space, a light thermally-produced gravitino 
is stable and can be a viable dark matter candidate. 
}
\arxivnumber{}
 
\begin{document}
\maketitle

%%%%%%%%%%%%%%%%%%%%%%%%%%%%%%%%%%%%%%%%%%%%%%%%%%%%%%%%%%%%%%%%%%%%%%%%%%%%%%%%%%%%%
%%%%%%%%%%%%%%%%%%%%%%%%%%%%%%%%%%%%%%%%%%%%%%%%%%%%%%%%%%%%%%%%%%%%%%%%%%%%%%%%%%%%%
%%%%%%%%%%%%%%%%%%%%%%%%%%%%%%%%%%%%%%%%%%%%%%%%%%%%%%%%%%%%%%%%%%%%%%%%%%%%%%%%%%%%%

%												Sec. 1 Introduction 

%%%%%%%%%%%%%%%%%%%%%%%%%%%%%%%%%%%%%%%%%%%%%%%%%%%%%%%%%%%%%%%%%%%%%%%%%%%%%%%%%%%%%
%%%%%%%%%%%%%%%%%%%%%%%%%%%%%%%%%%%%%%%%%%%%%%%%%%%%%%%%%%%%%%%%%%%%%%%%%%%%%%%%%%%%%
%%%%%%%%%%%%%%%%%%%%%%%%%%%%%%%%%%%%%%%%%%%%%%%%%%%%%%%%%%%%%%%%%%%%%%%%%%%%%%%%%%%%%

\section{Introduction}
Weak scale supersymmetry (SUSY) has long been the leading framework for addressing the hierarchy problem.  
However, after accumulating over 20 fb$^{-1}$ of data, the LHC has yet to find any evidence of superpartners near the TeV scale 
and has already placed tight constraints on the most compelling regions of SUSY parameter space.  
As the lower bounds on stop and higgsino masses approach the TeV range, there is generic tension with 
naturalness; at least some fine tuning is required to stabilize the electroweak scale. 

However, this interpretation of LHC results is model dependent since
 most SUSY searches assume $R$-parity conservation and, thus, require substantial MET in the final state. 
If this assumption is relaxed, sparticles can decay to standard model particles and the bounds 
become significantly weaker, thereby alleviating the tension with naturalness. 
  Since none of SUSY's theoretically desirable features strictly requires $R$-parity, the current experimental situation
  motivates serious efforts to construct viable $R$-parity violating (RPV) alternatives. 

\begin{figure}[t]
\begin{center}\includegraphics[width=9.5cm]{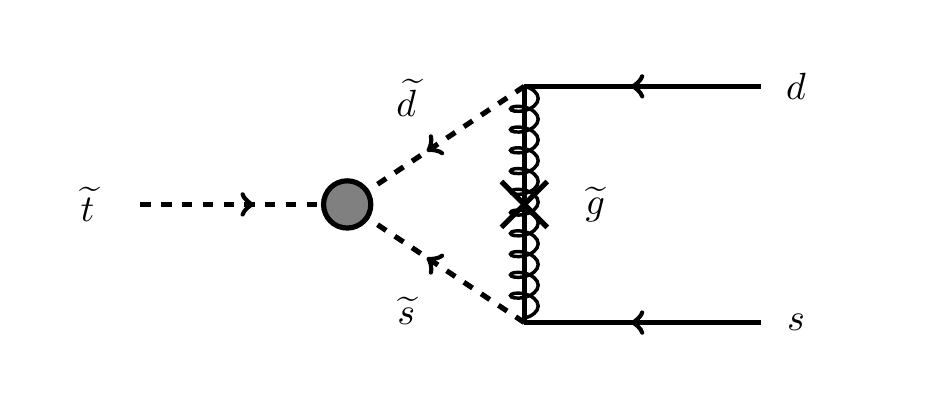}   \end{center}
\vspace{-0.6cm}         
\caption{The loop process that yields SUSY breaking fermion-scalar RPV interactions. }\label{fig:loopudd}
\end{figure}

In the absence of $R$-parity, the MSSM allows dangerous baryon and lepton violating operators in the superpotential
\be
W_{RPV} = \frac{\lambda_{ijk}}{2}   L_{i}L_{j}\bar E_{k} + \lambda^{\prime}_{ijk}  Q_{i}L_{j}\bar D_{k}  + \frac{\lambda^{\prime\prime}_{ijk} }{2} \,\bar U_{i} \bar D_{j} \bar D_{k} + \mu_{L_{i}} L_{i}H_{u} ~~~,
\label{eq:W-rpv}
\ee
and corresponding SUSY breaking terms in the soft Lagrangian 
\be \label{eq:soft-lag-intro}
{\cal L}_{\cancel{SUSY}} ~\supset  \frac{ {\cal A}_{ijk} }{2}\, \tilde L_{i} \tilde L_{j}  \tilde {\bar E}_{k} +  {\cal A}_{ijk}^{\prime} \, \tilde Q_{i} \tilde L_{j} \tilde {\bar D}_{k}   
+ \frac{     {\cal A}^{\prime \prime}_{ijk}        }{2} \, \tilde {\bar U}_{i} \tilde {\bar D}_{j}\tilde {\bar D}_{k} +  {\cal B}_{i} \tilde L_{i} H_{u}  + h.c. ~~, 
\ee
which induce rapid proton decay and unsuppressed FCNCs if the couplings in Eqs.\! (\ref{eq:W-rpv}) and (\ref {eq:soft-lag-intro}) are of natural size. 
Since proton decay typically requires both baryon and lepton number violation, 
the most stringent constraints can be evaded if leptonic RPV is strongly suppressed, but baryonic RPV via $\bar U\bar D
\bar D$ is large enough to allow the lightest squarks to decay promptly without MET \cite{Evans:2012bf, Brust:2012uf}. 
 
Several models in the literature satisfy these criteria. 
 Minimal Flavor Violating (MFV) SUSY \cite{Nikolidakis:2007fc, Csaki:2011uq}, for example, constrains
all flavor violating processes with the appropriate Yukawa couplings, which also determine the size and scope of allowed 
RPV interactions.  However, maintaining MFV structure in a UV complete scenario 
 requires nontrivial model building \cite{Krnjaic:2012aj, Franceschini:2013ne, Csaki:2013we}.       
Similarly, ``Collective RPV" \cite{Ruderman:2012jd} only allows RPV in particular
 combinations of couplings, so their overall effect yields the requisite suppression.  
 Other models with similar features are found in \cite{FileviezPerez:2011pt, Alves:2012fx, Bhattacherjee:2013gr, Allanach:2003eb, Steffen:2006hw, Florez:2013mxa}.
 
 Here we propose a novel scenario in which baryonic RPV arises at tree level in the
{\it soft terms}, but the scalar-fermion RPV interactions in Fig.~\ref{fig:loopudd} arise only at loop level with additional suppression.
These loop suppressed couplings can still be dangerous if RPV $\cal A$-terms are of 
order the weak scale.  
 For instance, if the baryon number violating ${\cal A}$-term (${\cal A}^{\prime \prime}$) is comparable to a typical
 soft mass $m_{_{\cal S}}$, 
\begin{equation}
 \lambda^{\prime\prime} \simeq\frac{g_s^2}{16\,\pi^2}\frac{{\cal A}^{\prime \prime}}{m_{ _{\cal S} }}\sim 10^{-2}~~~, 
\end{equation}
 this effective scalar-fermion coupling  
 is ruled out by precision flavor constraints, which require $\lambda^{\prime\prime} \lsim 10^{-7}$ for 
 light flavors \cite{Ellis:1984gi, Goity:uq, Allanach:1999ic}.
  However, if these terms are generated effectively through a
 heavy mediator of mass $M$ that ensures ${\cal A}^{\prime \prime} \sim m_{_{\cal S} }^2 / M$, then the amount of RPV is controlled dynamically.
In this framework, viable soft RPV can arise when $M$ is of order the SUSY breaking scale,
so no additional hierarchy is required. Although some aspects of soft RPV interactions have  been studied from a phenomenological perspective in
 \cite{BarShalom:2003cn, Carlos:fk,Jack:2004dv,Barbier:2004ez,Kord:2011jd}, to our knowledge, a realistic
 model has never been realized before.  
 
 Our model features a global $R$-symmetry that forbids RPV interactions in the superpotential. This symmetry
  is broken in a hidden sector and communicated to the MSSM through a heavy mediator that gets integrated out 
  to induce effective RPV $\cal A$-terms for squarks\footnote{A global $R$-symmetry can also yield purely leptonic RPV operators \cite{Frugiuele:2012pe} in the superpotential.}. 
  If gauge mediation communicates SUSY breaking to the visible sector, 
  the spectrum will also feature a metastable gravitino LSP that can be a viable dark matter candidate if thermally
  produced in the early universe.  

 The outline of this paper is as follows: in section \ref{sec:model}, we list the general criteria for soft RPV and present a concrete model based on gauge mediation; 
in section \ref{sec:constraints} we consider the experimental constraints and map out the allowed parameter space;
and in section \ref{sec:conclusion} we make some concluding remarks.

%%%%%%%%%%%%%%%%%%%%%%%%%%%%%%%%%%%%%%%%%%%%%%%%%%%%%%%%%%%%%%%%%%%%%%%%%%%%%%%%%%%%%
%%%%%%%%%%%%%%%%%%%%%%%%%%%%%%%%%%%%%%%%%%%%%%%%%%%%%%%%%%%%%%%%%%%%%%%%%%%%%%%%%%%%%
%%%%%%%%%%%%%%%%%%%%%%%%%%%%%%%%%%%%%%%%%%%%%%%%%%%%%%%%%%%%%%%%%%%%%%%%%%%%%%%%%%%%%

%												Sec 2. Model Setup 

%%%%%%%%%%%%%%%%%%%%%%%%%%%%%%%%%%%%%%%%%%%%%%%%%%%%%%%%%%%%%%%%%%%%%%%%%%%%%%%%%%%%%
%%%%%%%%%%%%%%%%%%%%%%%%%%%%%%%%%%%%%%%%%%%%%%%%%%%%%%%%%%%%%%%%%%%%%%%%%%%%%%%%%%%%%
%%%%%%%%%%%%%%%%%%%%%%%%%%%%%%%%%%%%%%%%%%%%%%%%%%%%%%%%%%%%%%%%%%%%%%%%%%%%%%%%%%%%%

\section{Model Description} \label{sec:model}
On general grounds, a viable model of soft RPV requires:
\begin{itemize}
\item Some symmetry $G$ that forbids the usual RPV interactions in the visible sector. 
\item A hidden sector (generically distinct from the SUSY breaking sector) 
that interacts with visible fields through a heavy mediator.
\item $G$-breaking triggered by soft terms in the hidden sector.
\end{itemize}
When the mediator is integrated out, the effective superpotential becomes
\be
W_{ef\!f} \supset \frac{X}{M}{\cal O}_{vis} + X F_{\displaystyle{\not}{G}}   ~~,
\ee
where $M$ is the heavy mediator mass, $X$ is a hidden sector superfield,
and $ F_{\displaystyle{\not}{G}}$ is a $G$ breaking spurion. The F-term for $X$
 induces a $G$-breaking ${\cal A}$-term $\sim  F_{\displaystyle{\not}{G}}/M$ 
for visible sector scalars,  while RPV interactions involving only visible fermions are
 forbidden at tree level when $\langle \tilde X \rangle =0$.

In this section we present a concrete model in which $G$ is an $R$-symmetry. 
To ensure predominantly baryonic RPV in the effective theory, we need 
lepton number to remain a good, accidental symmetry even after $R$-breaking.
 Fortunately this can be accomplished with an appropriate choice of hidden sector fields. 
 However, SUSY breaking typically contributes an additional source of $R$-breaking, so 
we need to ensure that the mediation mechanism doesn't spoil the accidental lepton symmetry.
Thus, we will use gauge mediation to communicate SUSY breaking to both visible and hidden sectors; 
perturbative gauge interactions preserve both lepton and baryon number, so 
leptonic RPV will not arise after $R$-breaking. 

%%%%%%%%%%%%%%%%%%%%%%%%%%%%%%%%%%%%%%%%%%%%%%%%%%%%%%%%%%%%%%%%%%%%%%%%%%%%%%%%%%%%%

%												2.1 Soft RPV from Gauge Mediation

%%%%%%%%%%%%%%%%%%%%%%%%%%%%%%%%%%%%%%%%%%%%%%%%%%%%%%%%%%%%%%%%%%%%%%%%%%%%%%%%%%%%%

\subsection{Soft RPV From a Broken $R$-symmetry} \label{ref:model-description}
\begin{figure}[t]
\begin{center}\includegraphics[width=12.5cm]{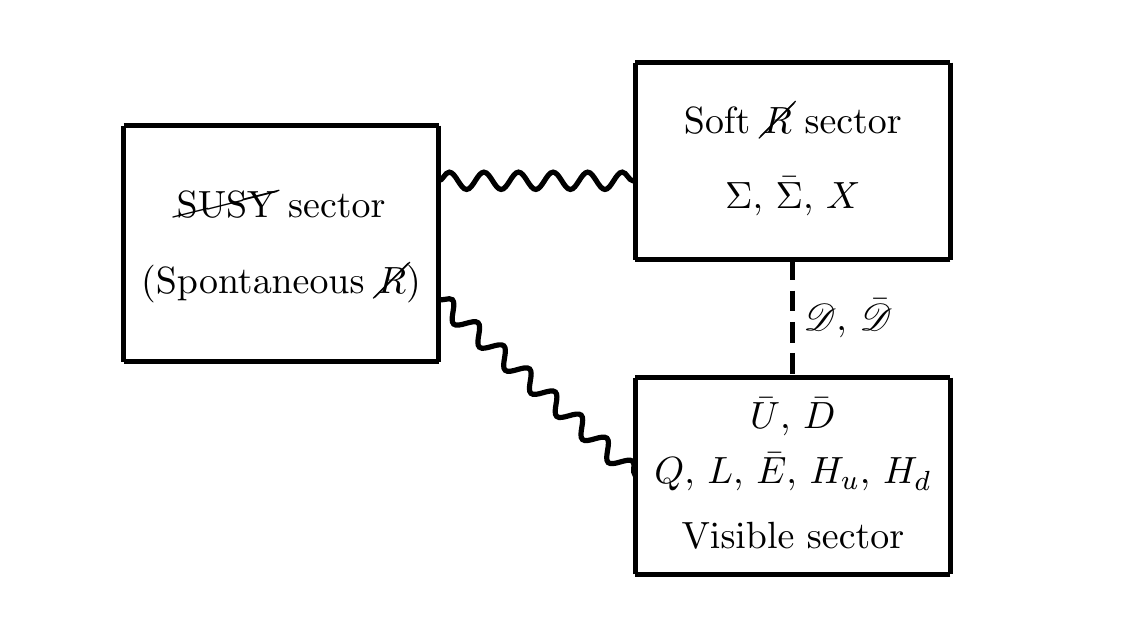}   \end{center}\vspace{-0.2cm}
\caption{A schematic diagram of the relevant sectors. SUSY breaking is communicated to both 
the $R$-breaking and visible sectors through gauge mediation. }
\label{fig:sectors}
\end{figure}
Since $R$-symmetries are vital for generic SUSY breaking \cite{Nelson:cq},  we begin 
by imposing the following $R$-charge assignments for MSSM fields 
\begin{equation} 
R[Q,\,\bar{U},\,\bar{D}]=1,\quad R[L]=4/3,\quad R[\bar E]=2/3,\quad R[H_u,\,H_d]=0 ~~,  \label{eq:r-charges}
\end{equation}
which forbid the RPV interactions in Eq.\,(\ref{eq:W-rpv}) without imposing $R$-parity. Although this choice of $R$-charges is anomalous, heavy spectators can 
 be added to cancel this anomaly without spoiling any of the model's features. 
The MSSM $\mu$ term is also forbidden at tree level, but one can arise if an additional singlet $S$ with $R$-charge +2 gets a VEV to induce $\langle S\rangle H_uH_d$ in 
the superpotential. It is also 
possible to generate weak scale higgsino and (Dirac) gaugino masses with an {\it unbroken} $R$-symmetry, though additional 
electroweak doublets are required \cite{Fox:2002bu, Kribs:2007ac}.
Since the novel features of our model do not depend on the 
details of the Higgs sector, we leave this issue for future work. 

The model contains three sectors depicted schematically in Fig.~\ref{fig:sectors}:
\begin{itemize}
\item {\bf Visible sector:} contains the usual MSSM fields and interactions consistent with the $R$-symmetry, which forbids
 RPV.  
\item {\bf SUSY breaking sector:} breaks both SUSY and the $R$-symmetry. SUSY 
breaking is mediated to the other sectors by gauge fields and decouples when all the gauge couplings vanish.
\item {\bf Soft $R$-breaking hidden sector:} features an additional $U(1)_H$ gauge symmetry so hidden scalars
get soft masses from gauge mediation. 
These soft masses can explicitly break the $R$-symmetry or induce radiative symmetry breaking through renormalization group evolution. 
$R$-breaking in this sector is communicated to the visible fields by heavy mediators $\mathscr D$ and $\bar {\mathscr D}$.
\end{itemize}
Even though the $R$-symmetry is also generically broken in the SUSY breaking sector, perturbative gauge interactions preserve both
 lepton and baryon number, so $R$-parity is not violated by gauge mediation. Visible sector
 RPV can only arise if the mediator connecting the visible and $R$-breaking sectors carries either lepton or 
 baryon number. In principle, the SUSY breaking and hidden sectors may be merged, but, for simplicity of exposition
 we ignore this possibility here. 

\begin{figure}[t]
\begin{center}
\begin{tabular*}{0.5\textwidth}{@{\extracolsep{\fill}}c|cccc}
	\hline
	\\[-7pt]
	 $\quad$ & $SU(3)_c$ & $U(1)_Y$ & $U(1)_H$ & $R$ \\[2pt]
	\hline\hline
	\\[-6pt]
	$\bar{U}$ & $\bar 3$ & $\!\!\!\!-2/3$ & $0$ &  $1$ \\[2pt]
	\\[-6pt]
	$\bar{D}$ & $\bar 3$ & $1/3$ & $0$ &  $1$\\[2pt]
	\hline\hline
	\\[-6pt]
	$\bar {\mathscr D}$ & $\bar 3$ & $1/3$ &  $0$ & $0$\\[2pt]
	\\[-6pt]
	$\mathscr D$ & $3$ & $\!\!\!\!-1/3$ & $0$ &  $2$\\[2pt]
	\hline\hline
	\\[-6pt]
	$X$ & $1$ & $0$ & $0$ &  $\!\!\!\!-1$\\[2pt]
	\\[-6pt]
	$\Sigma$ & $1$ & $0$ & $1$  & $3/2$\\[2pt]
	\\[-6pt]
	$\bar{\Sigma}$ & $1$ & $0$ & $\!\!\!\!-1$ & $3/2$\\[2pt]
	\hline
\end{tabular*}
\caption{The charge assignments in our model. From top to bottom: the right-handed quarks in the visible sector, the heavy mediators ${\mathscr D}\,\bar{{\mathscr D}}$, the singlet $X$ connects the mediators to the $\Sigma$ fields, which are charged under the gauged $U(1)_H$. The rightmost column lists $R$-charge assignments.}\label{table:particlecontent}
\end{center}
\end{figure}

\begin{figure}
	\begin{center} \hspace{-0.5cm}\includegraphics[width=15.5cm]{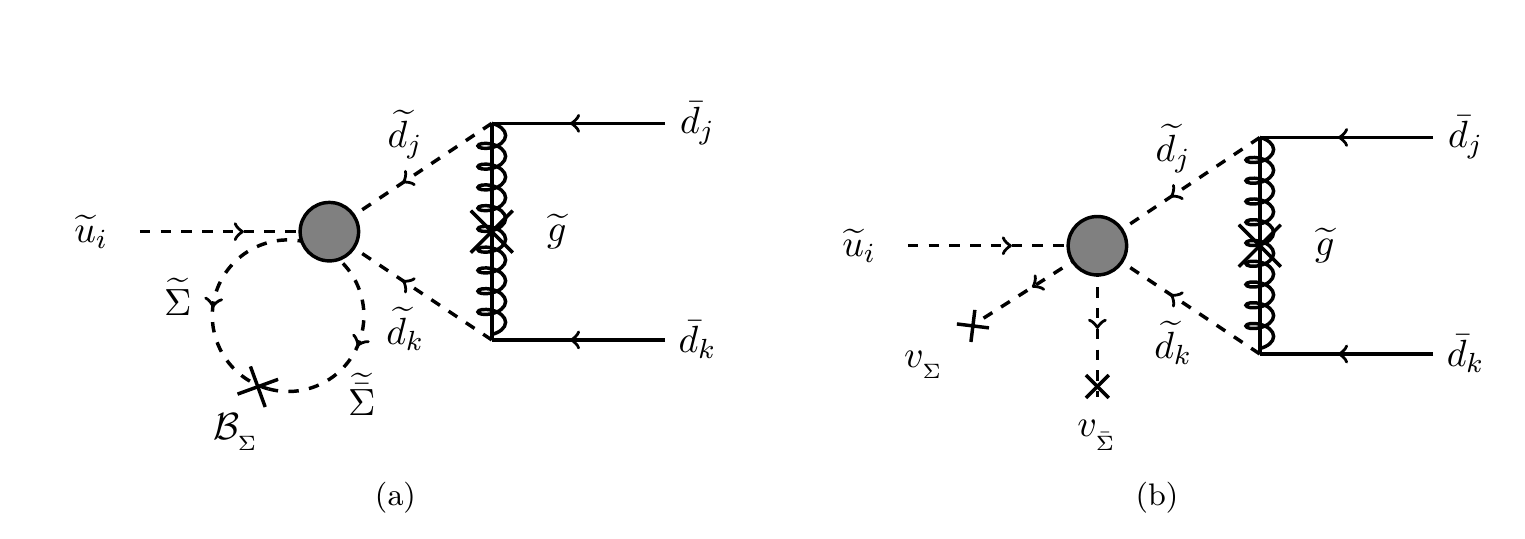}   \end{center}
	\vspace{-0.5cm}
	\caption{Effective $\lambda^{\prime\prime}$ couplings from a nonzero ${\cal B}$ term (a) and from
	spontaneous $R$-breaking (b). Diagrams
	with electroweak gauginos in place of gluinos also give subdominant contributions to this process. }          \label{fig:lambdas}
		\label{fig:loopSSB}
	\end{figure}

For the field content and charge assignments in Table~\ref{table:particlecontent},
 the most general, renormalizable superpotential for the new states is 
\begin{equation} \label{eq:superpot}
 \kappa_{ij}  \,\epsilon^{{abc}} \,\bar{U}_{a}^{i}\bar{D}^{j}_b\bar{\mathscr D}_{c}+ \kappa_i^\prime \bar{D}^{i} {\mathscr D} X +  \eta \, \Sigma\,\bar{\Sigma} X +  M_{_{\mathscr D}}\bar {\mathscr D} {\mathscr D} ~~~, 
\end{equation}
where $a,b,c$ are color indices and $i,j$ are flavor indices.
For $M_{\mathscr D} \gg m_{_{\cal S}}$, the heavy mediators ${\mathscr D}$ and $\bar{\mathscr D}$ are integrated out and the effective superpotential becomes 
\begin{equation} \label{eq:superpot}
 -\frac{ \kappa_{i[j}  \kappa_{k]}^\prime }{M_{_{\mathscr D}}}  \,\epsilon^{{abc}} \,\bar{U}^{i}_a\bar{D}^{j}_b\bar{D}^{k}_c X+\eta \,\Sigma\,\bar{\Sigma} X~~~,
\end{equation}
where the $j$ and $k$ indices are antisymmetrized.
If the scalar component of $X$ gets a vacuum expectation value (VEV), there will be baryonic RPV in both the soft terms and in the effective superpotential. 
To emphasize the novel features of this model, we assume $\langle \tilde X\rangle=0$ without 
essential loss of generality; we revisit this assumption in section \ref{sec:sigmavev}.
The effective scalar potential now contains 
\begin{equation} \label{eq:Fx}
|F_{X}|^2\supset    - \frac{  \kappa_{i[j}\kappa_{k]}^{\prime}\eta^*  }{M_{_{\mathscr D}}    } (\tilde{\Sigma}\,\tilde{\bar{\Sigma}})^*\, \tilde{\bar{U}^{i}} \tilde{\bar{D}^{j}}\tilde{\bar{D}^{k}} +c.c.~~~,
\end{equation}
and baryonic RPV arises from a $\tilde{\Sigma}$ and $\tilde{\bar{\Sigma}}$ loop with a $\cal B$-term (${\cal B}_{_{\Sigma}}$) insertion in Fig.~\ref{fig:lambdas}(a) or 
from $\Sigma$ and $\bar \Sigma$ VEVs ($v_{_{\Sigma}}$), which generate the diagram in Fig.~\ref{fig:lambdas}(b). Note that the 
$R$-charges in Eq.~(\ref{eq:r-charges}) are chosen to forbid the baryon and lepton number violating 
interaction $QL\bar {\mathscr D}$, which generates  $QL\bar D$ when the mediator is integrated out.\footnote{ Since gravity violates all global and discrete symmetries, Planck suppressed operators -- e.g. $\frac{1}{M_{pl}}QQQL$ and 
$\frac{1}{M_{pl}} \bar U\bar U\bar D\bar E$ --  
can still be dangerous if their coefficients are not suppressed \cite{Csaki:2013we}. As in the $R$-parity conserving MSSM, we assume these to be negligible or absent in a full theory
valid at the Planck scale.  }

Since gauge mediation communicates SUSY breaking to both visible and hidden sectors, the essential features  
of this model are insensitive
to the details of SUSY breaking and the field content of the messenger sector. These details will, however, 
 determine the relative sizes of ${\cal B}_{_{\Sigma}}$ and  $v_{_{\Sigma}}$, so for the remainder of this paper 
 we will remain agnostic about which diagram in Fig.~\ref{fig:lambdas} dominates and consider only
 the limiting cases in which only ${\cal B}_{_{\Sigma}}$ or $v_{_\Sigma}$ is nonzero.  
 The general case with both contributions merely interpolates between these extremes, so our 
 approach loses no essential generality.

%%%%%%%%%%%%%%%%%%%%%%%%%%%%%%%%%%%%%%%%%%%%%%%%%%%%%%%%%%%%%%%%%%%%%%%%%%%%%%%

%											2.2            Audd from a B-term

%%%%%%%%%%%%%%%%%%%%%%%%%%%%%%%%%%%%%%%%%%%%%%%%%%%%%%%%%%%%%%%%%%%%%%%%%%%%%%%

\subsection{$\cal B$-term $R$-breaking}\label{sec:bterm}
As a warmup to see the essential features of the model, we first consider a toy situation
 in which all $R$-breaking arises from a nonzero $\Sigma \bar\Sigma$ $\cal B$-term, but
the $U(1)_H$ remains unbroken. The   
$\tilde\Sigma$ and $\tilde{\bar \Sigma}$ scalars get positive soft masses ($m_{_\Sigma}$) from gauge mediation
so $v_{_{\Sigma}} =  v_{_{\bar \Sigma}}  = 0$, visible sector RPV arises from the effective ${\cal A}$ term 
\begin{equation}
{\cal A}_{ijk}^{\prime\prime}\simeq\frac{ \kappa_{i[j}\kappa_{k]}^{\prime}  \eta^*}{16\,\pi^2}\,\frac{{\cal B}_{_{\Sigma}}}{M_{_{\mathscr D}}}\,\log\frac{         M_{*}^2    }{   m^2_{_\Sigma}}~~,
\end{equation}  
where $M_{*}$ is the messenger scale, so the diagram in Fig.~\ref{fig:lambdas}(a) yields
\begin{equation}\label{eq:lambdaudd1}
\lambda^{\prime\prime}_{ijk}\simeq\frac{       \kappa_{i[j}\kappa_{k]}^{\prime}    \eta^*          g_s^2   \,{\cal B}_{_{\Sigma}}
      }{(16\,\pi^2)^2     \, M_{_{\mathscr D}}M_{\tilde{g}}            }\,
\,\log\frac{   M_{*}^2  }{m^2_{_\Sigma}}~~~.
\end{equation}
In section \ref{sec:constraints}, we will see that, for order one 
$\kappa, \kappa^\prime$, and $\eta$, and benchmark inputs
 $M_{_{\mathscr D}}\sim 10^4$ TeV, $M_{*}\sim 10^{9}$ TeV, $\sqrt{{\cal B}_{_{\Sigma}}}\sim M_{\tilde{g}}\sim m_{_\Sigma}\sim 1$ TeV, 
 the baryonic RPV coupling 
 $\lambda^{\prime\prime}$ is naturally of order $10^{-7}$ and safe from flavor constraints.

\begin{figure}[t]
\begin{center}\includegraphics[width=9.cm]{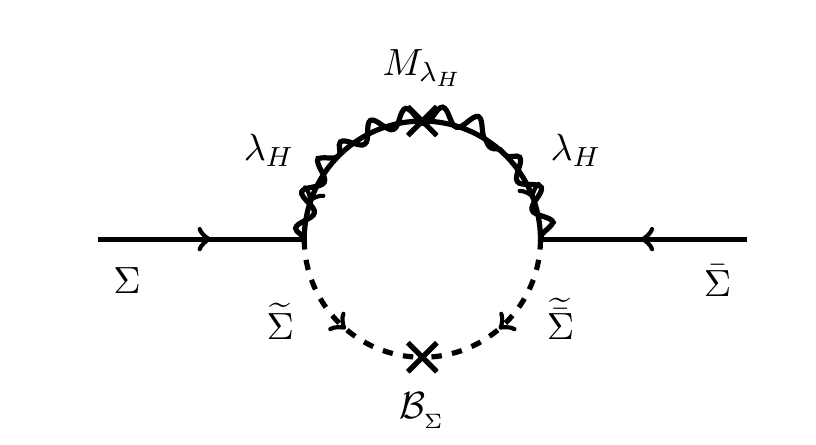}   \end{center}
	\vspace{-0.3cm}          
	\caption{ Dirac mass $\mu_{_\Sigma}$ for $\Sigma$ and  $\bar \Sigma$ from a nonzero ${\cal B}_{_{\Sigma}}$ term.} 
	\label{fig:bterm-mu}
\end{figure}

Although fermion mass terms for $\Sigma, \bar \Sigma$ and $X$ are forbidden at tree level, 
a Dirac mass $\mu_{_\Sigma}$ arises from hidden gaugino ($\lambda_H$) interactions 
 at one loop in Fig.~\ref{fig:bterm-mu},
\begin{equation}
\mu_{_\Sigma} \simeq\frac{g_{_H}^2}{16\,\pi^2}\frac{{\cal B}_{_{\Sigma}}}{\, M_{\lambda_H}} ~~~ ,
\label{eq:bmass}
\end{equation}
where $M_{\lambda_H} \sim m_{_\Sigma}$ is the hidden gaugino mass. An $X$ fermion mass $\sim \mu_\Sigma / 16\pi^2$ also 
arises with additional loop suppression from a similar diagram with $\Sigma, \bar \Sigma \to X$, $\lambda_H \to \Sigma$ and
$M_{\lambda_H} \to \mu_\Sigma$. 
In this phase, the dark gauge symmetry is unbroken, so the stable $\Sigma$ fermions annihilate to dark radiation in the early universe.
The $X$ fermions decay promptly through the $X \bar U \bar D \bar D$ operator so long as they are heavier than the proton. If 
 they are lighter than the $\sim 10$ MeV gravitino dark matter candidate  (see section \ref{sec:cosmology}), they can contribute to the
dark matter abundance without overclosing the universe. 

For $v_{_{\Sigma}} = 0$ in this minimal setup, the $X$ scalar is massless at tree level and acquires a tachyonic mass from loops of
 $\Sigma$ and $\bar\Sigma$ fermions. The resulting VEV generates potentially large superpotential RPV via
 $\langle X \rangle \bar U \bar D \bar D$, so this toy scenario is unstable unless $\tilde X$ acquires mass by other means. Additional
 mass terms for $X$ can arise either in the superpotential with additional $R$-charged fields or after SUSY breaking if the mediation
 mechanism gives gauge-singlets soft masses. 

%%%%%%%%%%%%%%%%%%%%%%%%%%%%%%%%%%%%%%%%%%%%%%%%%%%%%%%%%%%%%%%%%%%%%%%%%%%%%%%

%											2.2            Audd from Sigma VEV

%%%%%%%%%%%%%%%%%%%%%%%%%%%%%%%%%%%%%%%%%%%%%%%%%%%%%%%%%%%%%%%%%%%%%%%%%%%%%%%

\subsection{Spontaneous $R$-breaking}\label{sec:sigmavev}
Now we present a more concrete 
%\draftnote{can we say ``concrete"? sounds like the $B$-term cannot be realistic} 
scenario that generates soft RPV and solves this problem with nonzero VEVs $v_{_{\Sigma, \bar \Sigma}}$ that 
break both $U(1)_H$ and the $R$-symmetry. For simplicity
 we assume all hidden sector ${\cal A}$ and ${\cal B}$ terms vanish and
set $m_{_\Sigma} = m_{_{\bar\Sigma}}$, so the scalar potential contains
\begin{equation}\label{eq:sigmapot} \frac{g_{_H}^2}{2}\left(|\tilde\Sigma |^2-|\tilde{\bar{\Sigma}}|^2\right)^2  
+\eta^2\left(      |\tilde\Sigma\,\tilde{\bar{\Sigma}}|^2 +  |\tilde X\,\tilde\Sigma |^2+|\tilde X\,\tilde{\bar{\Sigma}}|^2     \right)     
-m^2_{_\Sigma}\left(|\tilde\Sigma|^2+|\tilde{\bar{\Sigma}}|^2 \right)~~,
\end{equation}
where the negative mass squared can arise through RG evolution if $\Sigma$ and $\bar\Sigma$ couple
to other fields with nonzero soft masses -- see Appendix \ref{Sec:vevappendix} for a concrete example. 

\begin{figure}[t]
\begin{center}\includegraphics[width=9.5cm]{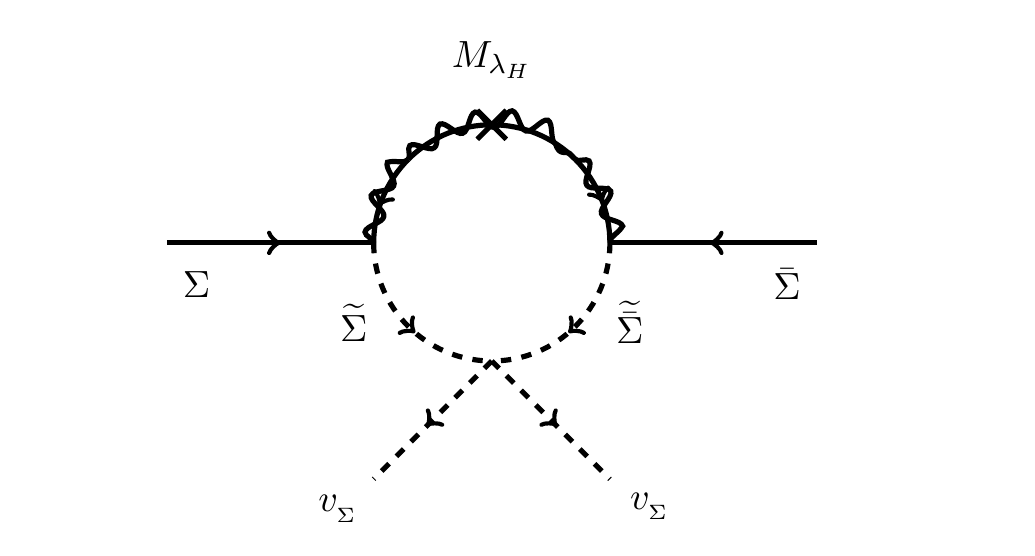}   \end{center}
	\vspace{-0.5cm}
	\caption{ 
	Loop level Dirac mass $\mu_{_\Sigma}$ for $\Sigma$ and $\bar \Sigma$ for $v_{_\Sigma} \ne 0$.} 
	\label{fig:fermion-mu}
\end{figure}

For $g_{_H}^2>\eta^2/2$, the classical minimum is 
\begin{equation}
v_{_\Sigma}=  v_{_{\bar \Sigma}}  = m_{_\Sigma}/\eta\qquad,\qquad \langle \tilde X\rangle=0 ~~~,
\end{equation}
but, quantum corrections still generate an $\tilde X$ VEV. However, unlike in section \ref{sec:bterm},
$\tilde X$ now has a tree level mass of  $m_X \sim v_{_\Sigma}$, so minimizing the Coleman-Weinberg potential yields 
$\langle \tilde X\rangle\propto \mu_{\Sigma}^3/m_X^2$, where
\begin{equation}
\mu_{_\Sigma} \simeq\frac{g_{_X}^2}{16\,\pi^2}\frac{\eta^2v_{_\Sigma}^2}{M_{\lambda_H}}  ~~ ,
\end{equation}
 is the $\Sigma\bar\Sigma$ Dirac mass that arises from the 
 %\draftnote{loop-suppressed}
 loop-diagram in Fig.~\ref{fig:fermion-mu}. 
 Thus, the $\langle X \rangle \bar U \bar D \bar D$ correction to fermionic 
RPV is subdominant to the soft contribution in Fig.~\ref{fig:loopSSB}(b) for which 
\begin{equation}\label{eq:lambdaudd2}
\lambda^{\prime\prime}_{ijk} = \frac{    \kappa_{i[j}\kappa_{k]}^{\prime} \eta^*\,       g_s^2  \,   v_{_{\Sigma}}^2       }{32\,\pi^2   \, M_{_{\mathscr D}}M_{\tilde{g}}   }\,\frac{}{}~~~.
\end{equation}
Since the effective potential also contains 
\begin{equation}
|F_{X}|^2=\Big|  \eta\,\tilde\Sigma\,\tilde{\bar{\Sigma}}            -\frac{      \kappa\,\kappa^{\prime}      }{       {M_{_{\mathscr D}}}    }      
    \tilde{\bar{U}}\tilde{\bar{D}}\tilde{\bar{D}}          \Big|^2 ~~ ,
\end{equation}
a nonzero $v_{_\Sigma}$ can, in principe, trigger color breaking, however,
 in Appendix~\ref{Sec:appendix1} we find that color remains unbroken so long as 
$v_{_\Sigma}\lsim M_{\mathscr D}\,\left(m_{\tilde{q}}/M_{\mathscr D}\right)^{3/4}$, 
where $m_{\tilde q}$ is a typical squark mass of order the weak scale.

The $R$-symmetry forbids superpotential mass terms for  $X$, 
$ \Sigma$ and ${\bar\Sigma}$, so the hidden 
 sector spectrum is entirely determined by SUSY breaking parameters. As in section \ref{sec:bterm}, 
  $\tilde \Sigma$ and $\tilde{ \bar \Sigma}$ get gauge mediated soft masses and $\tilde X$ gets a soft mass at one loop.
  After symmetry breaking,  the $X, \Sigma,\bar \Sigma$, and $\lambda_H$ fermions 
 mix and the resulting mass eigenstates are of order the electroweak scale. For generic
 mixing angles, all hidden sector mass eigenstates will be linear combinations of all four
interaction eigenstates, so they all decay promptly through the $X\bar U \bar D\bar D$ portal. 

A spontaneously broken $R$-symmetry gives rise to a massless $R$-axion that can accelerate supernova cooling and cause cosmological problems \cite{Raffelt:1990yz}.
 Conventionally, $R$-breaking arises only in the SUSY breaking sector and the BPR mechanism \cite{Bagger:1994hh} 
 generates an $R$-axion mass from a constant term in the superpotential introduced to cancel the cosmological constant. In
this scenario, our hidden sector also contributes to $R$-breaking, so the physical $R$-axion is now a linear combination of SUSY breaking 
and hidden sector states, but still acquires a BPR mass, so we will not consider it further. Although the BPR term explicitly breaks the 
$R$-symmetry, we assume its existence has no additional bearing on the symmetries of our superpotential; it serves merely as a placeholder
for the cosmological constant problem, which is beyond the scope of this work.

%%%%%%%%%%%%%%%%%%%%%%%%%%%%%%%%%%%%%%%%%%%%%%%%%%%%%%%%%%%%%%%%%%%%%%%%%%%%%%%%%%%%%
%%%%%%%%%%%%%%%%%%%%%%%%%%%%%%%%%%%%%%%%%%%%%%%%%%%%%%%%%%%%%%%%%%%%%%%%%%%%%%%%%%%%%
%%%%%%%%%%%%%%%%%%%%%%%%%%%%%%%%%%%%%%%%%%%%%%%%%%%%%%%%%%%%%%%%%%%%%%%%%%%%%%%%%%%%%

%												3. Experimental Bounds 

%%%%%%%%%%%%%%%%%%%%%%%%%%%%%%%%%%%%%%%%%%%%%%%%%%%%%%%%%%%%%%%%%%%%%%%%%%%%%%%%%%%%%
%%%%%%%%%%%%%%%%%%%%%%%%%%%%%%%%%%%%%%%%%%%%%%%%%%%%%%%%%%%%%%%%%%%%%%%%%%%%%%%%%%%%%
%%%%%%%%%%%%%%%%%%%%%%%%%%%%%%%%%%%%%%%%%%%%%%%%%%%%%%%%%%%%%%%%%%%%%%%%%%%%%%%%%%%%%

\section{Experimental Bounds}\label{sec:constraints}

  \begin{figure}[t!]
\begin{center}\includegraphics[width=8.cm]{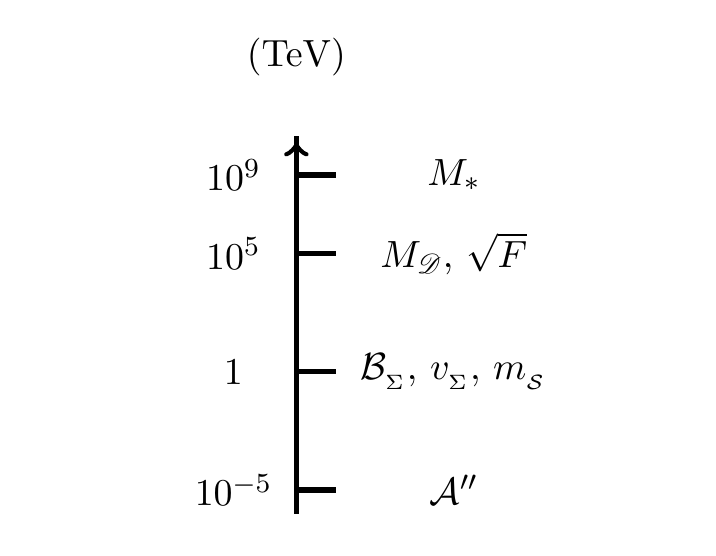}   \end{center} \vspace{-0.4cm}
\caption{The hierarchies of scales in our model. Since ${\cal A}^{\prime \prime} \sim {\cal B}_{_{\Sigma}} / M_{\mathscr D}\sim v_{_{\Sigma}}^2 / M_{\mathscr D}$  and $ M_{\mathscr D} \sim  \sqrt{F}$, 
this setup introduces no energy scales beyond those already required in conventional SUSY models. 
 }\label{fig:massscales}
\end{figure} 

\begin{figure}[t]
\begin{center}  \hspace{-1.13cm} \includegraphics[width=8cm]{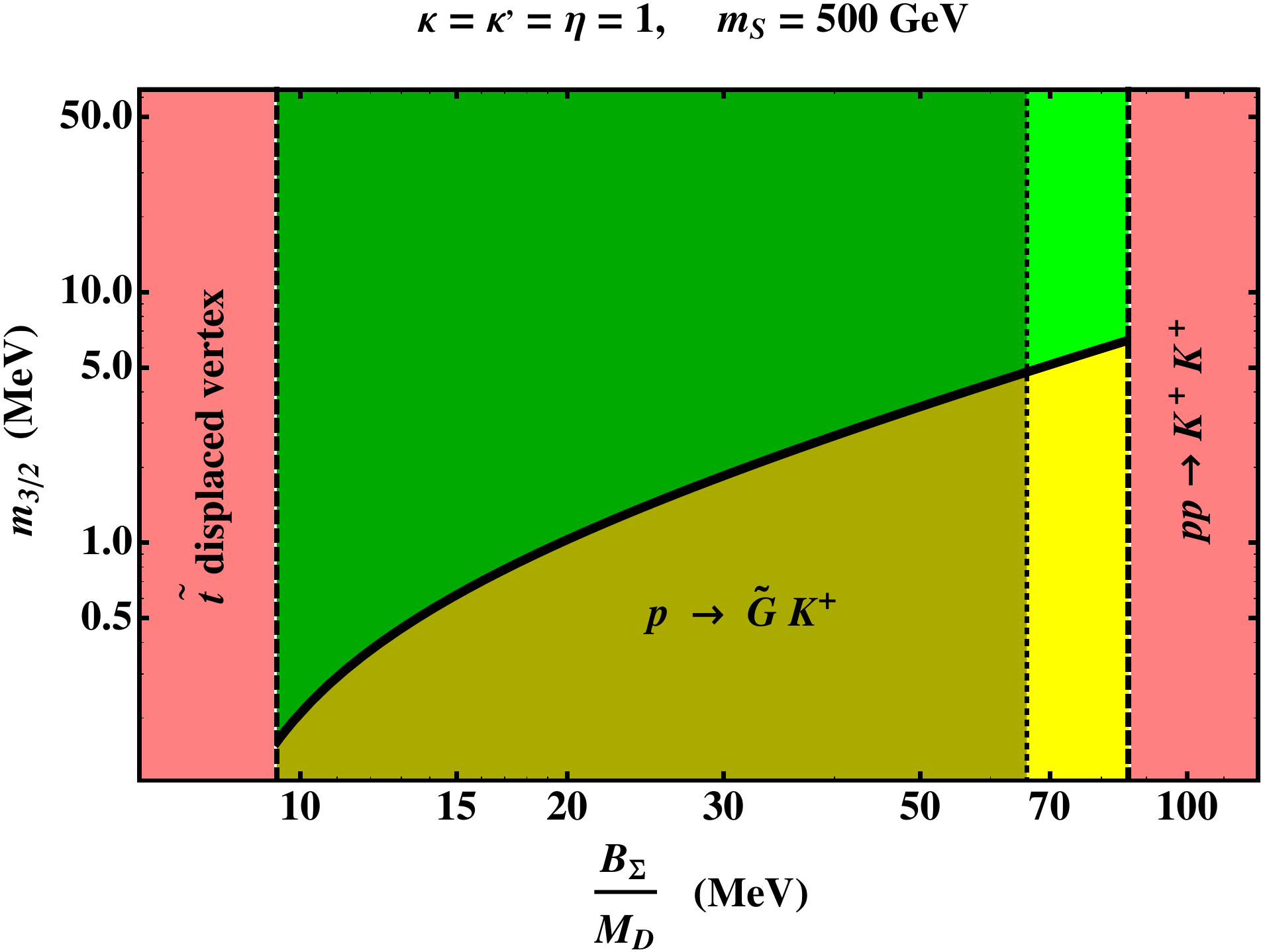}\hspace{0.cm}
\includegraphics[width=8cm]{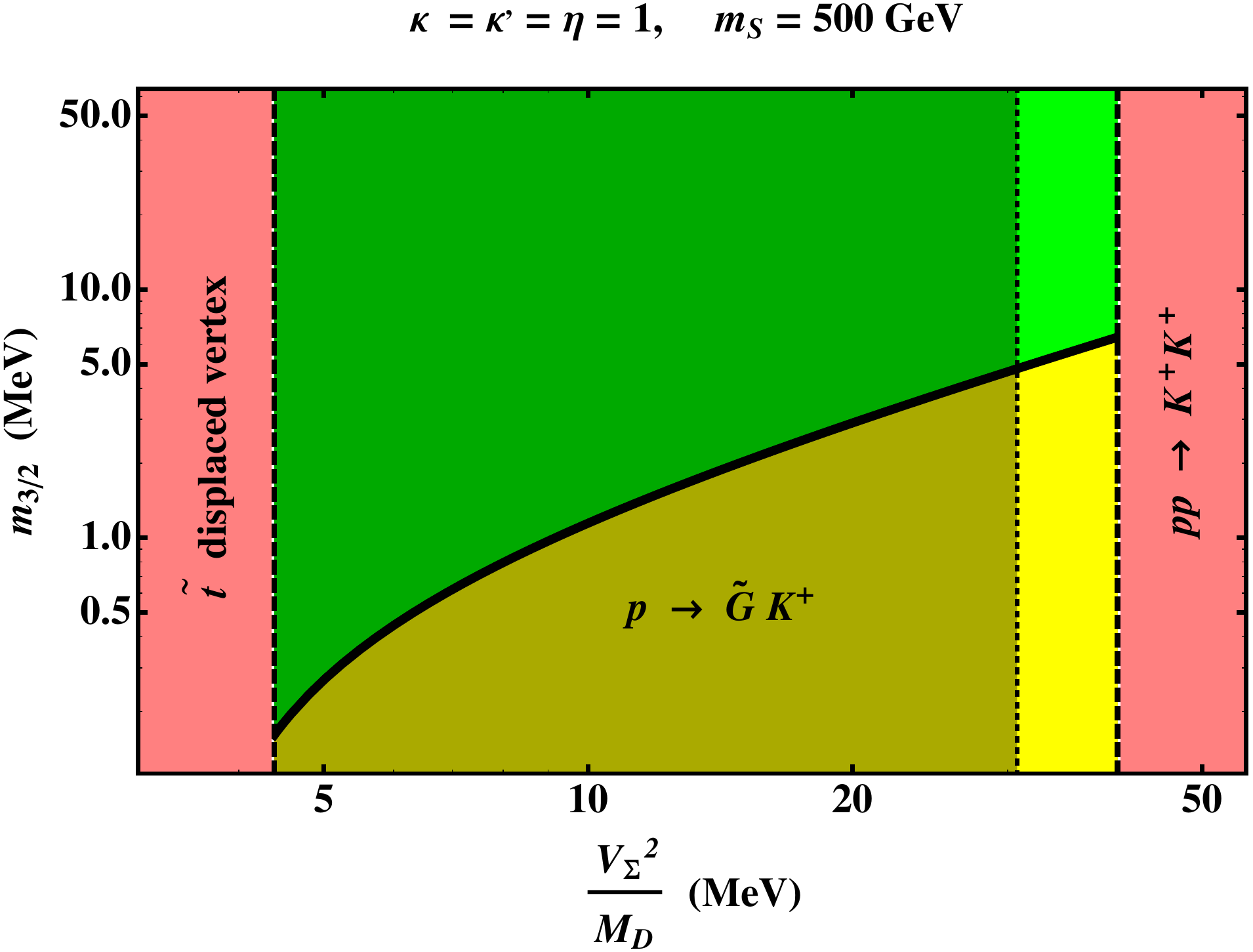}\end{center}\vspace{-0.4cm}
\caption{The parameter space for our model in the ${\cal B}$-term scenario (left) and in the spontaneously broken phase (right). In each case, the 
light (dark) green represents the allowed region where the stop decays with vertices smaller than 2 mm (10 cm). Here we assume the most conservative
 scenario with $|\kappa_{ij}| = |\kappa^{\prime}_{i}| = |\eta| =  1$ for all coefficients. The rates that determine the red excluded regions are quadratically sensitive to 
 these parameters, so if light flavors have smaller coefficients, the parameter space expands considerably. 
}\label{fig:paramspace}
\end{figure}

In this section we consider the experimental constraints on our realization of soft RPV. For simplicity, we will follow
 the organization of section \ref{ref:model-description} and separately constrain the cases in which 
  the {\cal B}-term  and $\Sigma, \bar \Sigma$ VEVs are solely responsible for $R$-breaking; the most general case
  interpolates between these extremes. The ladder of scales in Fig.~\ref{fig:massscales} summarizes the relative sizes of 
  various inputs in our model and the plots in Fig. \ref{fig:paramspace} carve out the allowed parameter space in both 
  ${\cal B}$-term and spontaneously broken scenarios.

 \subsection{Direct Production}\label{sec:production}
Although the parameter space for RPV spectra with sparticles below a TeV has recently been reduced, the sensitivity of these
bounds is driven primarily by lepton number violating processes. For purely baryonic RPV, the bounds are considerably weaker and 
can accommodate natural stops with $\sim 100$ GeV masses, provided they decay predominantly to dijets via $\bar U \bar D\bar D$ \cite{Evans:2012bf}.
For RPV gluinos decaying exclusively to $\tilde g \to t\tilde t$, the strongest experimental bound
 is now $\gsim 670$ GeV \cite{Aad:2011zb,ATLAS:2012dp, Curtin:2012rm, Cohen:2012yc, Asano:2012gj, Han:2012cu}, however, recasting 
 $R$-parity conserving SUSY searches may place a stronger $\sim 800$ GeV bound on the gluino mass \cite{Berger:2013sir}. 

\subsection{Baryon Number Violation}\label{sec:deltab} 

The $\bar U \bar D \bar D$ interaction explicitly violates baryon number, so our model faces constraints from the null results of 
several low energy searches. The strongest limits come from the bounds on the
characteristic timescales for dinucleon decay $(pp \to K^{+} K^{+})$ \cite{Litos:2010zra} 
and neutron-antineutron oscillation $(n-\bar n)$ \cite{Abe:2011ky}
\be
\tau_{pp \to KK} \ge 1.7 \times 10^{32} {\rm \, yrs.}~~, ~~~ \tau_{n-\bar n} &\ge& 2.44 \times 10^{8}  {\rm \, sec. } ~~~,
\ee
and from proton decay via $p \to K^{+} \nu$, for which the bound is \cite{Beringer:1900zz}  
\be
\tau_{p \to K^{+} \nu} \ge 2.3\times10^{33} {\, \rm yrs.} ~~.
\ee
Although our model doesn't violate lepton number, this bound conservatively constrains the $p\to K^{+} \tilde G$ decay, which has similar kinematics
for a sufficiently light gravitino. 

%%%%%%%%%%%%%%%%%%%%%%%%%%%%%%%%%%%%%%%%%%%%%%%%%%%%%%%%%%%%%%%%%%%%%%%%%%%%%%%%%%%%%%
% 											                Dinucleon Decay
%%%%%%%%%%%%%%%%%%%%%%%%%%%%%%%%%%%%%%%%%%%%%%%%%%%%%%%%%%%%%%%%%%%%%%%%%%%%%%%%%%%%%%

\subsubsection{Dinucleon Decay}\label{sec:dinucleon}
\begin{figure}[t]
\begin{center}\includegraphics[width=13cm]{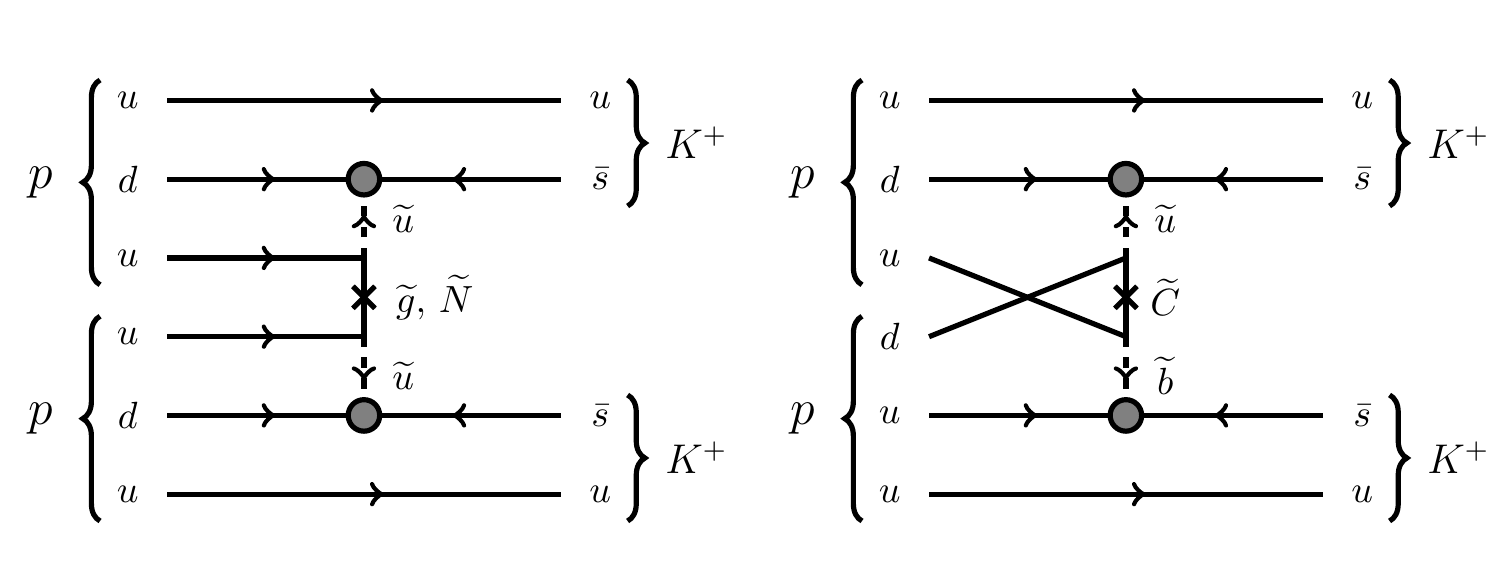}   \end{center}\vspace{-0.3cm}
	\caption{Dinucleon decay via baryonic RPV interactions. In the text we assume gluino exchange (left) dominates. }
	\label{fig:nndecay}
	\end{figure}

 Following Goity and Sher \cite{Goity:uq}, 
the dinucleon decay rate for the dominant processes shown in Fig.~\ref{fig:nndecay} is 
\begin{equation}
\Gamma_{pp\to KK}\sim\rho_N\,\frac{128\,\pi\,\alpha_{s}^2\,{\Lambda}^{10}}{m_p^2 \,m_{\tilde{u}}^8 \,M_{\tilde{g}}^2}\,\left(\lambda^{\prime\prime}_{uds}\right)^2 ~~~,
\end{equation}
where $m_{\tilde u}$ is the lightest up-type squark mass, $\rho_N\sim 0.25/{\rm \,fm}^3$ is the density of nuclear matter and $\Lambda$ is the characteristic hadronic energy-scale. 
Here we assume $M_{\tilde{C}}>M_{\tilde{g}}\,\alpha/\alpha_s\gsim 220$ GeV,
so the gluino exchange diagram in Fig.~\ref{fig:nndecay} dominates. 
Thus, satisfying the experimental bound $\tau_{pp\to KK}\geq 1.7\times 10^{32}$ yrs.  requires 
\begin{equation}\label{eq:nndecaybound}
\lambda^{\prime \prime}_{uds}~ \lsim ~2.5\times 10^{-7}\,\left(\frac{150\,{\rm MeV}}{\Lambda}\right)^{5/2}\,\left(\frac{M_{\tilde{g}}}{800\,{\rm GeV}}\right)^{1/2}\left(\frac{m_{\tilde{u}}}{500\,{\rm GeV}}\right)^2 ~~,
\end{equation}
Translating this into a constraint on the ${\cal B}$-term scenario ($v_{_{\Sigma}} = 0$) in section \ref{sec:bterm}, we have
\begin{equation}\label{eq:softmassbound3}
\frac{{\cal B}_{_{\Sigma}}}{M_{\mathscr D}} ~\lsim ~81\,{\rm MeV}\left(\frac{150\,{\rm MeV}}{\Lambda}\right)^{5/2}\,\left(\frac{M_{\tilde{g}}}{800\,{\rm GeV}}\right)^{1/2}\left(\frac{m_{\tilde{u}}}{500\,{\rm GeV}}\right)^2\,( \eta^{*} \kappa_{u[d} \kappa^{\prime}_{s]})^{-1} ~~ ,
\end{equation}
where we have set $M_{*} = 10^{9}$ GeV and $m_{_{\Sigma}} = 500$ GeV inside the log of Eq. (\ref{eq:lambdaudd1}).
Similarly, for the spontaneous $R$-breaking scenario (${\cal B}_{_{\Sigma}} = 0$) in section \ref{sec:sigmavev}, the corresponding bound is extracted
from Eq. (\ref{eq:lambdaudd2})   
\begin{equation}\label{eq:softmassbound4}
\frac{ v_{_{\Sigma}}^{2}    }{M_{\mathscr D}}~ \lsim ~ 42 \, {\rm MeV} \,\left(\frac{150\,{\rm MeV}}{\Lambda}\right)^{5/2}\,\left(\frac{M_{\tilde{g}}}{800\,{\rm GeV}}\right)^{1/2}\left(\frac{m_{\tilde{u}}}{500\,{\rm GeV}}\right)^2\,( \eta^{*} \kappa_{u[d} \kappa^{\prime}_{s]})^{-1} ~~ .
\end{equation}     
Unlike similar processes in MFV SUSY \cite{Csaki:2011uq} where the light quark couplings are Yukawa suppressed, our setup
imposes no necessary hierarchies in the RPV couplings. 

\subsubsection{$n-\bar n$ Oscillation}\label{sec:nnosc}
Unlike dinucleon decay, $n-\bar{n}$ oscillation also requires 
flavor violation from $R$-parity conserving vertices.
However, aside from the baryon violating ${\cal A}$-term, all visible sector soft masses arise directly from gauge mediation, so their 
 flavor structure comes entirely from Yukawa couplings. Thus, up to an overall coefficient, our $n-\bar n$ oscillation amplitudes are identical 
 to those computed in \cite{Csaki:2011uq}.

Chirality-preserving flavor-violating masses arise predominantly from
 MSSM $F$-terms after SUSY and electroweak symmetry breaking through 
\begin{equation}
\tilde{Q}^\dagger\left( v_{u}^{2}\,Y_uY_u^{\dagger} + v_{d}^{2}\,Y_dY_d^{\dagger}  \right)\tilde{Q} ~~~,
\end{equation}
and similar terms for $\tilde{\bar U}$ and $\tilde{ \bar D}$, where $Y_{u,d}$ are Yukawa matrices. For simplicity, we take the Higgs 
doublet VEVs $v_{u,d}$ to be at the soft scale $\sim m_{_{\cal S}}$. 
In gauge mediation, chirality flipping $\cal A$-terms arise only at higher order and suffer both Yukawa and loop suppression, so they are typically smaller than soft masses. 
 However, different realizations of gauge mediation give rise to $\cal A$ terms
 with different degrees of suppression relative to the soft scale. Since we remain agnostic about the details of the messenger sector, 
 we conservatively parametrize any possible suppression with the general ansatz ${\cal A} \equiv \epsilon m_{_{\cal S}}$.
 
\begin{figure}[t]
\begin{center}\includegraphics[width=10.cm]{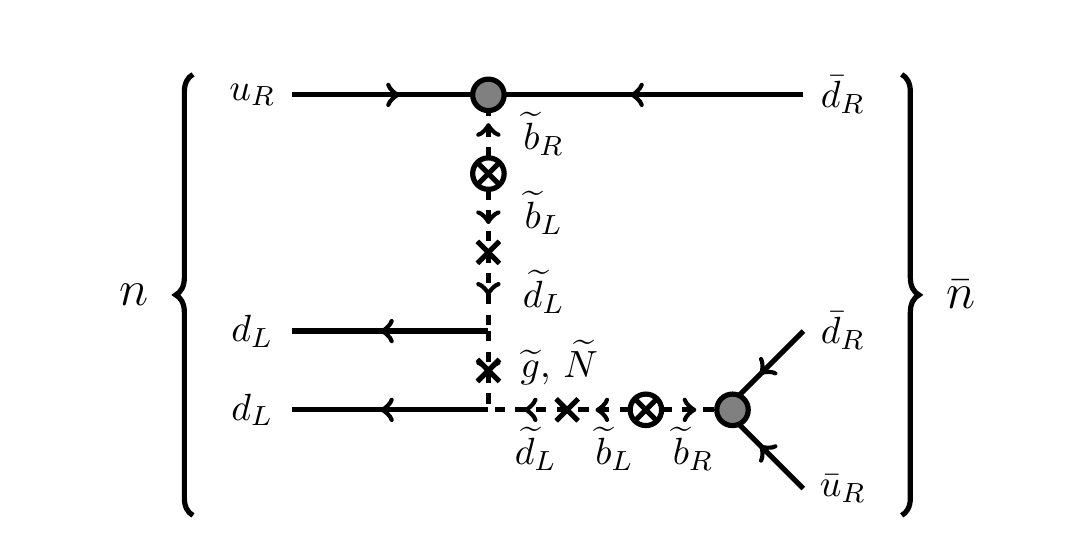}   \end{center}
\vspace{-0.4cm}
\caption{The dominant diagram for neutron anti-neutron oscillation. All $R$-parity conserving soft masses and $\cal A$-terms are consistent with MFV. }\label{fig:nnbar}
\end{figure}

Putting all the squarks at a common soft mass $m_{\tilde q} \sim m_{_{\cal S}}$, the amplitude for the dominant diagram shown in Fig.~\ref{fig:nnbar} is 
\begin{equation}
\mathcal{M}_{n-\bar{n}}\sim  \, g^2_s \epsilon^{2} \lambda^6\,  \Lambda \left(\frac{  \Lambda }{m_{\tilde{q}}}\right)^4 \! \left(\frac{\Lambda}{M_{\tilde{g}}}\right)({ \lambda^{\prime\prime}_{udb} })^{ 2}  ~~ ,
\end{equation}
where $\lambda \simeq 0.23$ comes from the approximate CKM matrix parametrization in \cite{Csaki:2011uq}. 
The oscillation timescale is approximately $\tau_{n-\bar n}\sim\mathcal{M}^{-1}$, thus the experimental bound $\tau_{n-\bar n}\geq 2.44\times 10^8$ sec. requires 
\begin{equation}
\lambda^{\prime\prime}_{udb}\lsim 1.7\times10^{-6}\, \epsilon^{-2}\,\left(\frac{m_{\tilde{q}}}{500\,{\rm GeV}}\right)^4\,\left(\frac{250\,\rm{MeV}}{{\Lambda}}\right)^6\,\left(\frac{M_{\tilde{g}}}{800\,\rm{GeV}}\right) ~~ ,
\end{equation}
which is weaker than the bound from dinucleon decay in Eq.~\ref{eq:nndecaybound} even when $\epsilon$ is order one.

%%%%%%%%%%%%%%%%%%%%%%%%%%%%%%%%%%%%%%%%%%%%%%%%%%%%%%%%%%%%%%%%%%%%%%%%%%%%
%												3.2 Proton Decay
%%%%%%%%%%%%%%%%%%%%%%%%%%%%%%%%%%%%%%%%%%%%%%%%%%%%%%%%%%%%%%%%%%%%%%%%%%%%

\subsubsection{Proton Decay}\label{sec:protondecay}
\begin{figure}
\begin{center}\includegraphics[width=10.5cm]{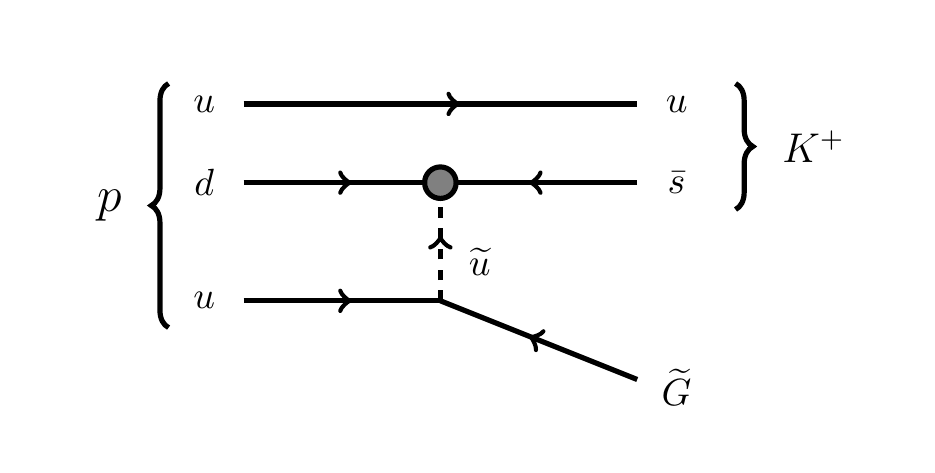}   \end{center}
\vspace{-0.7cm}
\caption{Proton decay via $p\to K^{+} \tilde G$.}\label{fg:pdecay}
\end{figure}

Since gauge-mediation typically features a light, sub-GeV gravitino, proton decay to $K^+ \tilde G$ through the diagram in Fig.~\ref{fg:pdecay}
may be kinematically allowed. The rate for this process is 
\begin{equation}
\Gamma_{p \to K^+\tilde G}~ \sim~\frac{m_p}{8\,\pi}\left(\frac{{\Lambda}}{m_{\tilde{u}}}\right)^4\left(\frac{\Lambda^2}{\sqrt{3}\,m_{3/2}\,M_{pl}}\right)^2\left(\lambda^{\prime\prime}_{uds}\right)^2~~,
\end{equation}
 and the lifetime for this channel must be longer than $2.3\times 10^{33}$ yrs., so the gravitino mass bound is 
\begin{equation}
m_{3/2}~\geq~ 4.7\,{\rm MeV} \left(\frac{\Lambda}{250\,{\rm MeV}}\right)^4\left(\frac{500\,{\rm GeV}}{m_{\tilde{u}}}\right)^2\left(\frac{\lambda^{\prime\prime}_{uds}}{10^{-7}}\right)~~,
\end{equation}
For $m_{3/2}\gsim 5$ MeV, this implies a lower bound on the SUSY breaking scale
\begin{equation}\label{eq:FMbound} 
\sqrt{{ F}}~\gsim~ 3.2\,\times 10^5\,{\rm TeV} ~~.
\end{equation}
If minimal gauge mediation gives rise to soft masses, the messenger scale $M_{*}$ must also 
 satisfy
 \be
M_* ~\gsim~1.3\times 10^{9}\,{\rm TeV}\,\left(\frac{500\,{\rm GeV}}{m_{_{\cal S}}}\right)~~.
 \ee
 
\subsection{Displaced Vertices} \label{sec:stopdecay}
To avoid MET searches at the LHC, sparticles must decay on collider timescales, so there is an upper bound on the lightest squark's lifetime.
 Although there are many LHC searches for displaced vertices \cite{Aad:2011zb,Asano:2011ri}, 
hadronically-decaying long-lived particles are significantly harder to constrain \cite{Graham:2012th}; 
viable decay lengths can even exceed $\sim 10 {\rm \, cm}$, so a dedicated search is necessary.
Given these uncertainties, we consider the experimental bounds in two 
regimes: for prompt decays, we conservatively require decay lengths $\ell_{\tilde q}<2$ mm; for signatures with viable displaced vertices, 
we demand $\ell_{\tilde q}<10$ cm, so most sparticles decay inside the tracker before reaching the hadronic calorimeter (HCAL), 
but may still be found with a dedicated search. 
 
 The width for a hardronically decaying stop NLSP\footnote{For typical SUSY breaking scales we consider, the gravitino is the LSP, though
for extremely high SUSY breaking scales, this need not be the case.} in its rest frame is 
\begin{equation}
\Gamma_{\tilde t \to \bar q\bar q}=\frac{m_{\tilde{t}}}{8\,\pi}\,\sin^2\theta_{\tilde{t}}\,|\lambda_{tqq}^{\prime\prime}|^2~~~,
\end{equation}
where $\theta_{\tilde t}$ is the stop mixing angle.  In the lab frame, the decay length is 
 $\ell_{\tilde t}\simeq \gamma\,\Gamma_{\tilde t \to q\bar q}^{-1}$, where $\gamma$ is the stop 
 boost factor; for a $300$ GeV stop and an $800$ GeV gluino produced at rest, $\gamma \sim 2$. For the remainder of this section
 we assume, for simplicity, that $ \gamma \sin^2\theta_{\tilde{t}}=1$. 
 
Assuming the dominant stop decay is $\tilde t \to \bar d \bar s$, the bound on $\lambda^{\prime\prime}_{tds}$ is 
\begin{equation}\label{eq:lambdalowerbound1}
\lambda^{\prime\prime}_{tds}>(0.26-1.8)\times 10^{-7}\left(\frac{300\,{\rm GeV}}{m_{\tilde{t}}}\right)^{1/2}.
\end{equation}
where the left and right numbers represent the bound assuming 10 cm and 2 mm displaced-vertex limits, respectively. 
For the $\cal B$-term scenario ($v_{_{\Sigma}} = 0, {\cal B}_{_{\Sigma}} \ne 0$) in section \ref{sec:bterm}, this implies
\begin{equation}\label{eq:BMbounlowerbound1}
\frac{{\cal B}_{_{\Sigma}}}{M_{{\mathscr D}}}~\gsim~ (8.3-58)\times \,{\rm MeV}\,\left(\frac{m_{\tilde{g}}}{800\,{\rm GeV}}\right)\left(\frac{300\,{\rm GeV}}{m_{\tilde{t}}}\right)^{1/2}\,
(\kappa_{t[d} \kappa^\prime_{s]} \eta^*)^{-1} ~~~, 
\end{equation}
with $m_{\Sigma}=1$ TeV and $M_*\sim 10^{9}$ TeV inside the log in Eq.~(\ref{eq:lambdaudd1}). Similarly, for the spontaneously broken scenario ($v_{_{\Sigma}} \ne 0, {\cal B}_{_{\Sigma}} = 0$) in section \ref{sec:sigmavev}, we have
\begin{equation}\label{eq:BMbounlowerbound2}
\frac{v_{_{\Sigma}}^2}{M_{{\mathscr D}}}~\gsim~ (4.3-31)\times \,{\rm MeV}\,\left(\frac{m_{\tilde{g}}}{800\,{\rm GeV}}\right)\left(\frac{300\,{\rm GeV}}{m_{\tilde{t}}}\right)^{1/2}\,(\kappa_{t[d} \kappa^\prime_{s]} \eta^*)^{-1}~~~.
\end{equation}
These bounds assume the stop is the lightest squark and decays predominantly through RPV interactions.
Thus, the only other kinematically allowed process $\tilde t \to t\,\tilde G$
must have a negligible branching ratio, which requires 
\begin{equation}
\Gamma_{\tilde t \to \bar q\bar q} \gg \Gamma_{t\,\tilde{G}}= \frac{m_{\tilde{t}}^5}{16\,\pi\,F^2}  ~~.
\end{equation}
As long as the SUSY breaking scale satisfies $\sqrt{F}>10^2$ TeV, the RPV branching ratio exceeds $99\%$. This constraint
 is trivially satisfied by considerations from proton decay in section \ref{sec:protondecay} above. 
 
%%%%%%%%%%%%%%%%%%%%%%%%%%%%%%%%%%%%%%%%%%%%%%%%%%%%%%%%%%%%%%%%%%%%%%%%%%%%
%												3.4 Dark Matter 
%%%%%%%%%%%%%%%%%%%%%%%%%%%%%%%%%%%%%%%%%%%%%%%%%%%%%%%%%%%%%%%%%%%%%%%%%%%%

\subsection{Gravitino Dark Matter} \label{sec:cosmology}
Since gauge mediation communicates SUSY breaking to the visible sector, the gravitino is the LSP with mass 
$m_{3/2} \sim {F}/M_{pl} \sim  {\cal O}(10)$ MeV for 
$\sqrt{F} \sim 10^8$ GeV. In this mass range $m_{3/2} < m_{p}$, so the process $\tilde G \to qqq$ is kinematically forbidden and the gravitino
is stable. Since sparticles rarely decay to gravitinos and their annihilation rate is suppressed by the SUSY breaking scale, the
present day  abundance is 
thermally generated \cite{Steffen:2006hw}
\be
\Omega_{3/2} h^{2} \simeq 0.1 \left(  \frac{T_{R}}{ 10^{5} {\rm\, GeV}}     \right)  \left(    \frac{m_{3/2}}{ 20 \, {\rm MeV} } \right)^{-1}  \left(    \frac{M_{\tilde g}}{ 800\, \rm GeV } \right)^{2} ~~ ,
\ee
where $T_{R}$ is the reheating temperature, so the RPV gravitino is a viable dark matter candidate.

%%%%%%%%%%%%%%%%%%%%%%%%%%%%%%%%%%%%%%%%%%%%%%%%%%%%%%%%%%%%%%%%%%%%%%%%%%%%
%%%%%%%%%%%%%%%%%%%%%%%%%%%%%%%%%%%%%%%%%%%%%%%%%%%%%%%%%%%%%%%%%%%%%%%%%%%%

%												4. Conclusion

%%%%%%%%%%%%%%%%%%%%%%%%%%%%%%%%%%%%%%%%%%%%%%%%%%%%%%%%%%%%%%%%%%%%%%%%%%%%
%%%%%%%%%%%%%%%%%%%%%%%%%%%%%%%%%%%%%%%%%%%%%%%%%%%%%%%%%%%%%%%%%%%%%%%%%%%%

\section{Conclusions}\label{sec:conclusion}

In this paper we have presented a new realization of weak scale SUSY with $R$-parity violation. 
Unlike conventional scenarios, suppressed baryonic RPV arises in the {\it soft terms}  
 when an $R$-symmetry is broken in a hidden sector and a heavy mediator is
integrated out; lepton number remains a good accidental symmetry. RPV interactions between quarks and squarks
arise at one loop and receive additional suppression. The model features 
light ($\sim$ few 100 GeV) squarks that decay promptly to hadrons and evade LHC searches in viable regions 
of parameter space safe from flavor constraints.  

For weak-scale $R$-breaking, the heavy mediator masses can be near the 
 SUSY breaking scale $\sqrt{F}\sim 10^{8}$ GeV to generate RPV couplings 
 with the requisite suppression, so the model requires no new scales beyond those already present in
   conventional SUSY models. If gauge mediation communicates
 SUSY breaking, the model also features a light $\sim 1- 100$ MeV gravitino 
 with a thermal abundance. For a reheating temperature of order $10^{5}$ GeV and a weak scale  
 gluino, a  gravitino in this mass range is a viable dark matter candidate. 
 However, gauge mediation serves merely as a 
 convenient mechanism to generate soft masses without violating lepton or baryon number; any
 alternative for which this holds true would work equally well. 
 
 If $R$-breaking arises from a ${\cal B}$-term for $\Sigma$ and $\bar\Sigma$ as in section \ref{sec:bterm}, the model 
requires either non-minimal gauge mediation to generate sizable ${\cal B}$-terms, or another
mediation mechanism that preserves the accidental lepton symmetry. We leave these model building
details for future work. For the more-concrete spontaneous $R$-breaking 
scenario in section \ref{sec:sigmavev}, the model requires either additional fields 
to drive radiative symmetry breaking for $\Sigma$ and $\bar \Sigma$ or an alternative to gauge mediation
that results in tachyonic soft masses in the hidden sector. In Appendix \ref{Sec:vevappendix} we show that 
radiative symmetry breaking is feasible, but leave other alternatives for future work. 

Grand unification with RPV is challenging because both lepton and baryon number violating
RPV interactions generally arise from the same interaction term. In $SU(5)$, for instance, $\bar U\bar D\bar D, QL\bar D$
and $LL\bar E$ all live in the same $10 \bar 5 \bar 5$ UV operator, so generating predominantly baryonic RPV at low energies 
requires additional model building gymnastics \cite{Bhattacherjee:2013gr}. In our case, the 
$R$-charge assignments differ for quark and lepton superfields, so it is not clear whether 
grand unification is possible.

%%%%%%%%%%%%%%%%%%%
%%%%%%%%%%%%%%%%%%%
\section*{Acknowledgments}
We thank Csaba Csaki, Ben Heidenreich, Markus Luty,  Surjeet Rajendran, Carlos Tamarit, John Terning and Yue Zhao for helpful conversations. 
Additional thanks to Ben Heidenreich, Carlos Tamarit, and John Terning for comments on the draft. Research at the Perimeter Institute is supported in part by the Government of Canada through Industry Canada, and by the Province of Ontario through the Ministry of Research and Information (MRI). 
YT thanks the Perimeter Institute for its hospitality while this work was in progress. YT is supported by the Department 
of Energy under grant DE-FG02-91ER406746.

%%%%%%%%%%%%%%%%%%%%%%%%%%%%%%%%%%%%%%%%%%%%%%%%%%%%%%%%%%%%%%%%%%%%%%%%%%%%
%%%%%%%%%%%%%%%%%%%%%%%%%%%%%%%%%%%%%%%%%%%%%%%%%%%%%%%%%%%%%%%%%%%%%%%%%%%%

%												A. VEV Appendix

%%%%%%%%%%%%%%%%%%%%%%%%%%%%%%%%%%%%%%%%%%%%%%%%%%%%%%%%%%%%%%%%%%%%%%%%%%%%
%%%%%%%%%%%%%%%%%%%%%%%%%%%%%%%%%%%%%%%%%%%%%%%%%%%%%%%%%%%%%%%%%%%%%%%%%%%%

\appendix
\section{Hidden Sector VEVs}\label{Sec:vevappendix}
%%%%%%%%%%%%%%%%%%%%%%%%%%%%%%%%
Throughout the paper, we have assumed that the $\Sigma$ and $\bar \Sigma$ scalars acquire negative mass-squared parameters 
to induce spontaneous symmetry breaking. Since the minimal superpotential only allows the $\Sigma X \bar \Sigma$ interaction and
 gauge mediation gives rise to positive soft masses, the setup requires either a nonminimal messenger sector to generate negative soft masses 
or substantial RG evolution. Since we are agnostic about the details of gauge mediation, here we present a concrete example of 
radiative $R$-breaking in the hidden sector as a proof of concept.  

If the $\Sigma$ scalars also couple to chiral fields
 $Y$ and $\bar Y$ with identical $R$-charges of $1/4$ and $U(1)_{H}$ charges of $\mp 1/2$, the superpotential also contains 
\be\label{eq:new-interactions}
W \supset \eta\,\Sigma\,X\,\bar{\Sigma}+\lambda_{ Y} \Sigma Y^{2} +  \lambda_{\bar Y} \bar \Sigma \bar Y^{2} ~~~,
 \ee
where $\eta$, $\lambda_{Y}$, and $\lambda_{Y}$ are order one parameters. Including $U(1)_{H}$ gauge interactions, the full set of  
RGEs is 
 \be
  &&\frac{dg_{_H}}{dt} =\frac{5 \,g_{_H}^3}{32\pi^2}  \label{eq;RG0} ~~~~~~~~
 \\
  &&\frac{d\lambda_{_{Y,\bar Y}}}{dt}=\frac{\lambda_{Y,\bar Y}}{16\pi^2} \left(\frac{5}{2}\lambda_{Y,\bar Y}^2-3g_{_H}^2\right)\label{eq;RG1}~~~~~~~~
  \\
  &&\frac{d\eta}{dt}=\frac{\eta}{16\pi^2} \left(3\eta^2-4g_{_H}^2\right)\label{eq;RG15}~~~~~~~~
 \\
  &&\frac{dm_{_\Sigma}^{2}}{dt}  =   \frac{1}{ 16\pi^{2}} \biggl[ \eta^{2} (2 m_{_\Sigma}^{2} + m_{_{\bar \Sigma}}^{2}+m_X^2)  + 4 \lambda_{Y}^{2} (m_{_\Sigma}^{2} + m_{Y}^{2})  
 + \frac{g_{_H}^{2}}{2}( - 2m_{_{\bar \Sigma}}^{2}+m_{\bar Y}^{2} -m_{Y}^{2} )  \biggr]  \label{eq;RG2} ~~~~~~~~
 \\ 
 &&\frac{dm_{_{\bar \Sigma}}^{2}}{dt} =   \frac{1}{ 16\pi^{2}} \biggl[ \eta^{2} (2 m_{_{\bar  \Sigma}}^{2} + m_{_\Sigma}^{2}+m_X^2)  + 4 \lambda_{Y}^{2} (\bar m_{_\Sigma}^{2} + \bar m_{Y}^{2})  
 + \frac{g_{_H}^{2}}{2}( - 2m_{_\Sigma}^{2}+m_{Y}^{2} -m_{\bar Y}^{2} )  \biggr]  \label{eq;RG3} ~~~~~~~~
 \\ 		  
 &&\frac{dm_{_{Y}}^{2}}{dt} =   \frac{1}{16 \pi^{2}} \biggl[  
  \lambda_{_Y}^{2} (4 m^{2}_{_\Sigma} + 6 m^{2}_{Y})     +   \frac{g_{_H}^{2}}{2} ( -m^{2}_{_\Sigma}+ m^{2}_{_{\bar \Sigma}}  - 
  \frac{1}{2} m^{2}_{_{\bar Y}})  \biggr]   \label{eq;RG4}  ~~~~~~~~  \\
&&\frac{dm_{_{\bar Y}}^{2}}{dt}  =   \frac{1}{ 16 \pi^{2}} \biggl[   \lambda_{\bar Y}^{2} (4 m^{2}_{_{\bar \Sigma}} + 6 m^{2}_{\bar Y})     +   \frac{g_{_H}^{2}}{2} ( -m^{2}_{_{\bar \Sigma}}  +  m^{2}_{_\Sigma}  -
  \frac{1}{2}m^{2}_{_Y})\biggr]     \label{eq;RG5} 
~~~~~~~~\\
  &&\frac{dm_{X}^{2}}{dt} =   \frac{1}{ 16 \pi^{2}} \biggl[   \eta^{2} (m^{2}_{_\Sigma} +  m^{2}_{_{\bar \Sigma}} + 2 m^{2}_{X}) \biggr]     \label{eq;RG6} ~~~~~~~~
  \ee
  Note that, without the interactions in Eq.~(\ref{eq:new-interactions}), the  
 $m^2_{\Sigma, \bar \Sigma}$ equations can be rewritten in terms of $ x \equiv m^2_{\Sigma}  +  m^2_{\bar \Sigma}$ so that 
 both become $dx/dt \propto x$ whose solution never runs negative.

\begin{figure}
\begin{center}\includegraphics[width=6cm]{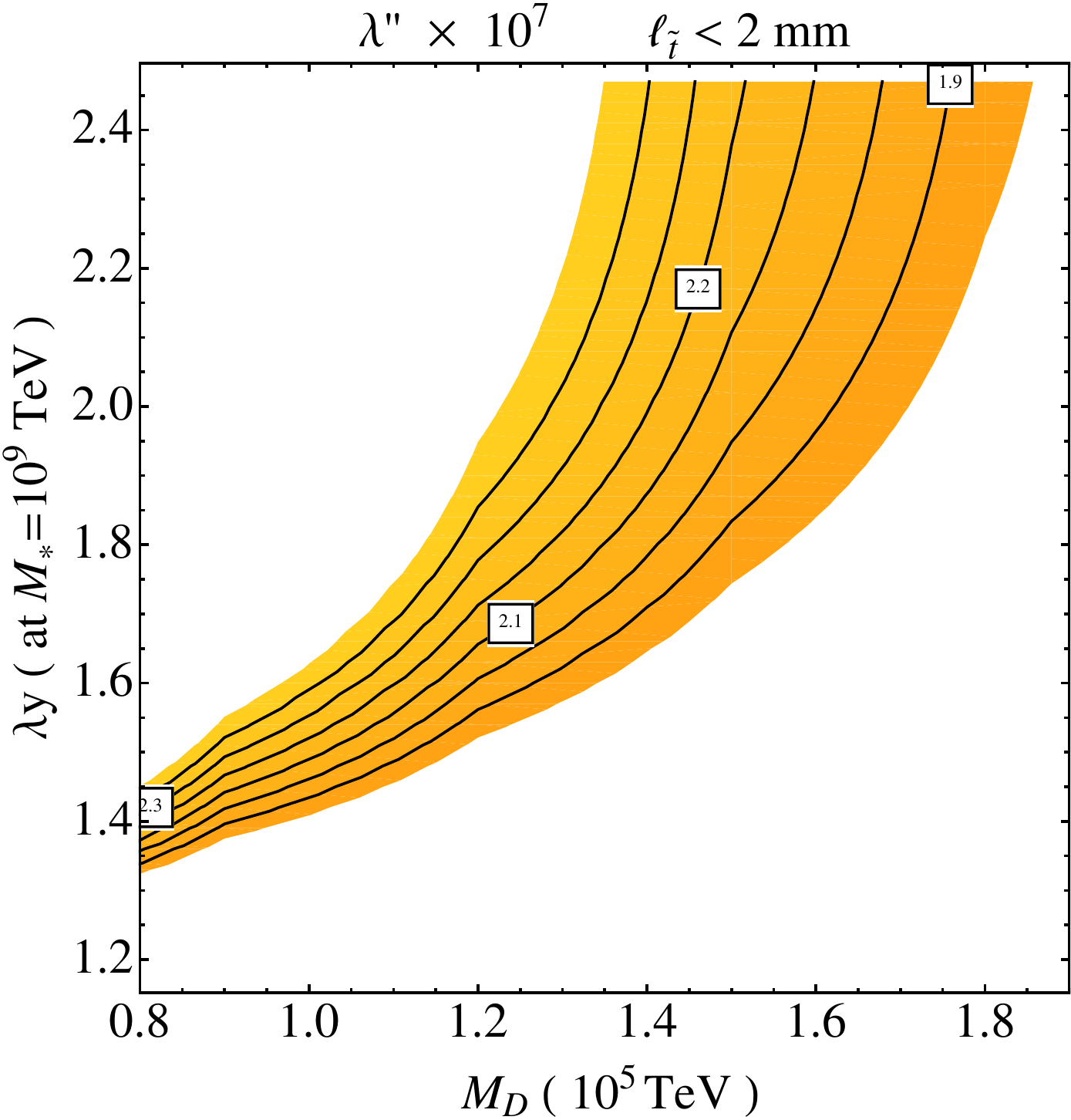}\,\,\,\,\,\,\,\,\,\,\,\,\,\,\,
\includegraphics[width=6cm]{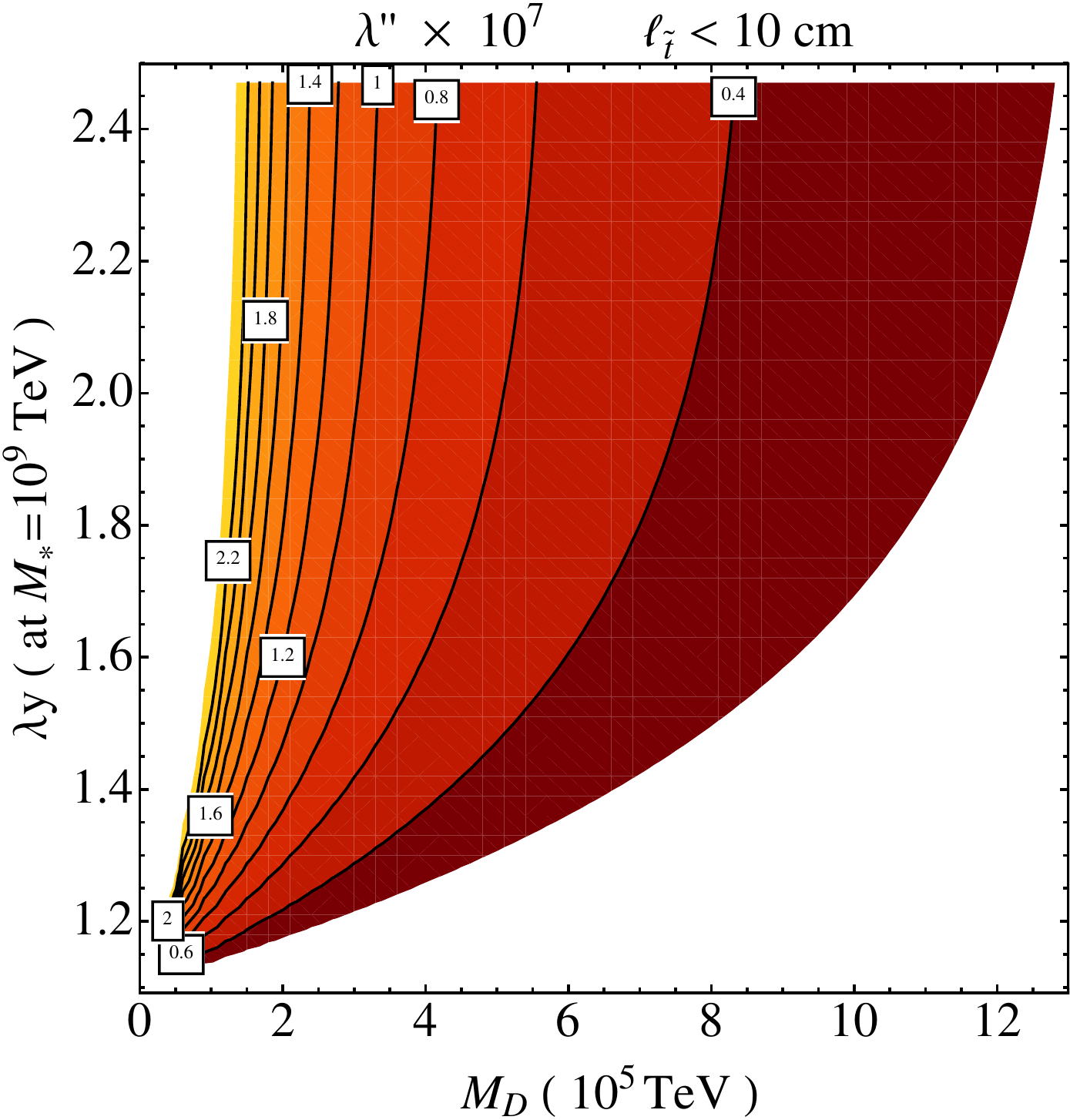}
\end{center} \vspace{-0.2cm}
\caption{
The allowed parameters space for $M_{{\mathscr D}}$ and $\lambda_Y$ with contours of $\lambda^{\prime \prime}$ derived from Eq. (2.7). The white region is excluded by 
the dinucleon decay and displaced vertex bounds in section 3. For the left plot, we assume prompt stop decays  with lengths $< 2$ mm; for
 the right plot we assume displaced stop decays inside the tracker ($< 10$ cm). The VEVs are computed after RG evolution with a  UV boundary
condition at  the messenger scale, $M_* = 10^9$ TeV, and IR boundary at the soft mass scale $m_{_{\cal S}} = 1$ TeV. We also assume flavor universal couplings $|\kappa|$ and $|\kappa^\prime|$ and soft 
masses dictated by gauge mediation. Note that the range of $M_{\mathscr D}$ is of order the benchmark SUSY breaking scale $\sqrt{F}\sim 10^5$ TeV.
}\label{fig:VEV-contour}
\end{figure}

In Fig.~\ref{fig:VEV-contour}, we plot contours of radiatively generated $\lambda^{\prime\prime}$ from Eq.~(\ref{eq:lambdaudd2})
 in the $M_{{\mathscr D}}, \lambda_{Y}$ plane. For each contour, minimal gauge mediation 
 defines the UV boundary condition $m_{_{\Sigma}}  = \frac{g_{_H}^2}{16\pi^2}\frac{F}{M_{*}}$
where $F$ saturates the bound in Eq.~(\ref{eq:FMbound}).  
 The allowed region assumes all couplings $\eta, \lambda_{_{Y,\bar Y}}, g_{_H}$ are all unity and we choose $\eta=0.1$ 
 at the EW scale to generate a larger $v_{\Sigma}$ and satisfy the bounds on
  $\lambda^{\prime\prime}$ from Eqs.~(\ref{eq:nndecaybound}) and ~(\ref{eq:lambdalowerbound1}).

For this field content, radiative symmetry breaking requires $Y$ $\bar Y$ to have larger soft masses than $\Sigma$ and $\bar \Sigma$
 at the mediation scale, which is not realized in the minimal minimal gauge mediation; $\Sigma$ and $\bar \Sigma$ have 
 larger gauge charges. However this can be accommodated if the $Y$ and $\bar Y$ carry additional gauge 
 charges to give them larger soft masses  at the mediation scale. Our example here assumes 
 $m_{Y,\bar Y}(M_{*}) = 2 m_{\Sigma, \bar \Sigma}(M_{*})$ and suffices 
to demonstrate that radiative $R$-breaking is possible.

%%%%%%%%%%%%%%%%%%%%%%%%%%%%%%%%%%%%%%%%%%%%%%%%%%%%%%%%%%%%%%%%%%%%%%%%%%%%
%%%%%%%%%%%%%%%%%%%%%%%%%%%%%%%%%%%%%%%%%%%%%%%%%%%%%%%%%%%%%%%%%%%%%%%%%%%%

%												B. "Color Breaking ?" Appendix

%%%%%%%%%%%%%%%%%%%%%%%%%%%%%%%%%%%%%%%%%%%%%%%%%%%%%%%%%%%%%%%%%%%%%%%%%%%%
%%%%%%%%%%%%%%%%%%%%%%%%%%%%%%%%%%%%%%%%%%%%%%%%%%%%%%%%%%%%%%%%%%%%%%%%%%%%

\section{Color Breaking?}\label{Sec:appendix1}
After $R$-breaking, up to order-one coefficients, the scalar potential derived from Eq.~(\ref{eq:superpot}) contains the terms 
\begin{equation}
V\supset \Big|\frac{\tilde{\bar{U}}\,\tilde{\bar{D}}_i\tilde{\bar{D}}_j}{M_{{\mathscr D}}}+v_{_\Sigma}^2\big|^2+m_{\tilde u}^2\,\big|\tilde{\bar{U}}\big|^2+m_{\tilde{d}_i}^2\,\big|\tilde{\bar{D}}_i\big|^2+m_{\tilde{d}_j}^2\,\big|\tilde{\bar{D}}_j\big|^2.
\end{equation}
 which can break color if squark masses are too small.
For simplicity, assuming  identical squark soft masses and positive superpotential couplings, we can rewrite the potential in
terms of dimensionless variables 
\begin{equation}
\hat{V}\equiv\frac{      m_{     \tilde{d}_i     }^2 \,       m_{\tilde{d}_j}^2      }{
m_{   \tilde{u}     }^4\,M_{{\mathscr D}}^4}\,V\supset \Big|x\,y\,z+s^2\Big|^2+\hat{m}^2\left(x^2+y^2+z^2\right)~~~,
\end{equation}
where 
\begin{equation}
x\equiv \frac{\langle\tilde{\bar{U}}\rangle}{M_{{\mathscr D}}},\quad y\equiv \frac{m_{\tilde{d}_i}\langle\tilde{\bar{D}}_i\rangle}{m_{\tilde{u}}\,M_{{\mathscr D}}},
\quad z\equiv \frac{m_{  \tilde{d}_j      }\langle\tilde{\bar{D}}_j\rangle}{m_{     \tilde{u}  }\,M_{{\mathscr D}}},\quad s\equiv \frac{\sqrt{m_{\tilde{d}_i}\,m_{\tilde{d}_j}     }\,}{m_{\tilde{u}}\,M_{{\mathscr D}}}  v_{_\Sigma}  ,\quad \hat{m}\equiv \frac{m_{\tilde{d}_i}     \,          m_{\tilde{d}_j}        }{      m_{\tilde{u}}      \,    M_{{\mathscr D}}       }~~~.
\end{equation}
At their extremal values, $x = y = z$, we demand
%By symmetry  we can set $x = y = z$ and demand   
\begin{equation}
(x^3+s^2)^2+3\,\hat{m}^2x^2\geq s^4~~~,
\end{equation}
to avoid color breaking at the global minimum. This conditions implies, $s\lsim \hat{m}^{3/4}$, so we need 
\begin{equation}
v_{\Sigma}\lsim\left(\frac{m_{\tilde{u}} \,m_{\tilde{d}_i}\,m_{\tilde{d}_j}}{M_{{\mathscr D}}^3}\right)^{1/4}\,M_{{\mathscr D}} ~~~~.
\end{equation}
For the model's relevant parameter space, $m_{\tilde d} \gsim 500$ GeV and $M\simeq 10^5$ TeV, this constraint 
becomes $v_{\Sigma}^2/M_{{\mathscr D}}\lsim 10^{-4}$ TeV, which is an order of magnitude weaker than the dinucleon decay bound  in Eq.~(\ref{eq:softmassbound4}),
 so color remains unbroken for the viable parameter space we consider.

%%%%%%%%%%%%%%%%%%%
%%%%%%%%%%%%%%%%%%%

% Bibliography
\bibliographystyle{utphys}
\bibliography{./SoftRPV}

\end{document}